\documentclass[12pt,dvips -P pdf -t letter -o Reflectivitiy.ps Reflectivity.dvi]{iopart}

\newcommand{\lsmo}{LaSrMnO$_{4}$}
\usepackage[english]{babel}
\usepackage[T1]{fontenc}
\usepackage{iopams} 
\usepackage{graphicx}
\usepackage{epstopdf}
\usepackage{cite}
\usepackage{color}
\usepackage{soul}

\begin{document}

\title[]{Electronic depth profiles with atomic layer resolution from resonant soft x-ray reflectivity}

\author{M. Zwiebler$^1$, J. E. Hamann-Borrero$^1$, M. Vafaee$^2$, P. Komissinskiy$^2$, S. Macke$^{3,4}$, R. Sutarto$^5$, F. He$^5$, B. B\"uchner$^1$, G. A. Sawatzky$^6$, L. Alff$^2$ and J. Geck$^1$}

\address{$^1$Leibniz Institute for Solid State and Materials Research, IFW-Dresden, Helmholtzstr. 20, 01069 Dresden, Germany}
\address{$^2$Institute of Materials Science, Technische Universit\"at Darmstadt, 64287 Darmstadt, Germany}
\address{$^3$Max Planck-UBC Centre for Quantum Materials, Vancouver, V6T1Z1, Canada}
\address{$^4$Max Planck Institute for Solid State Research, Heisenbergstr. 1, 70569 Stuttgart, Germany}
\address{$^5$Canadian Light Source, University of Saskatchewan, Saskatoon, Saskatchewan, S7N0X4, Canada}
\address{$^6$Department of Physics and Astronomy, University of British Columbia, Vancouver, V6T1Z1, Canada}

\begin{abstract}
The analysis of x-ray reflectivity data from artificial heterostructures usually relies on the homogeneity of optical properties of the constituent materials. However, when the x-ray energy is tuned to an absorption edge, this homogeneity no longer exists. Within the same material, spatial regions containing elements at resonance will have optical properties very different from regions without resonating sites. In this situation, models assuming homogeneous optical properties throughout the material can fail to describe the reflectivity adequately. As we show here, resonant soft x-ray reflectivity is sensitive to these variations, even though the wavelength is typically large as compared to the atomic distances over which the optical properties vary. We have therefore developed a scheme for analyzing resonant soft x-ray reflectivity data, which takes the atomic structure of a material into account by "slicing" it into atomic planes with characteristic optical properties. Using \lsmo\/ as an example, we discuss both the theoretical and experimental implications of this approach. Our analysis not only allows to determine important structural information such as interface terminations and stacking of atomic layers, but also enables to extract depth-resolved spectroscopic information with atomic resolution, thus enhancing the capability of the technique to study emergent phenomena at surfaces and interfaces.
\end{abstract}


\section{Introduction}
Specular x-ray reflectivity is one of the work horses for characterizing thin films and multilayers. In simple words, the reflectivity is given by interference of x-rays that are reflected at the different interfaces realized in such an artificial heterostructure. Referring to the reflection of optical light, an interface can be defined as a region in space where there is a change of the refractive index $n$. Similarly, also in the x-ray range even a small change in $n$ will introduce an interface, thus a traveling x-ray wave will be reflected. This high interface sensitivity is what allows to accurately determine structural properties of heterostructures such as layer thicknesses and interface roughnesses by means of  x-ray reflectivity.

Recently, with the advent of synchrotron radiation, the availability of photon sources with very high brilliance and tunable energy has opened the frontiers for x-ray reflectivity techniques to study additional properties apart from structure. Electronic properties, for instance, can be studied by tuning the x-ray photon energies to an absorption edge. At these so-called resonant energies, the refractive index depends very strongly on the valence shell properties of the resonant scattering centers and hence, the sensitivity to spatial variations of the electronic properties is dramatically enhanced. This renders resonant x-ray reflectivity (RXR) an ideal tool to study electronic properties and phenomena at surfaces and buried interfaces in an element specific and non-destructive way.

The development of RXR was in particular triggered by the recent progress made in the atomic scale synthesis of transition metal oxide (TMO) heterostructures. TMOs provide perhaps one of the richest and fruitful fields in condensed matter research in terms of electronic properties and emerging novel physics \cite{Dagotto07,zubko11,Chakhalian12,hwang12}. Examples of these exotic phenomena are, among others, the formation of a two-dimensional electron gas at the polar/non-polar interfaces of LaAlO$_3$/SrTiO$_3$ \cite{ohtomo04} or the proximity effects and orbital reconstruction in superconductor/ferromagnet (SC/FM) interfaces \cite{Buzdin05,Chakhalian06,Chakhalian07}. All these properties are closely related to the transition metal (TM) $3d$ and oxygen $2p$ electrons and their interaction with the crystal lattice.

RXR experiments have therefore in particular been performed at the TM $L_{2,3}$ edges, where the $3d$ electrons of the TM are directly probed. In this way, important information has been obtained e.g. about the spatial electron density redistribution of the Ni $3d$ electrons in LaNiO$_3$/LaAlO$_3$ multilayers \cite{eva11} or the Co valence reconstruction at a LaCoO$_3$ polar film surface \cite{hamann14}. Moreover, employing the x-ray magnetic circular dichroism (XMCD) effect, the magnetization profile of SC/FM interfaces \cite{brueck11}, exchanged bias systems \cite{Tonnerre08,Brueck10} and other multilayers \cite{Brown08,tonnerre12,Jal13,Hosoito14} have been studied.

The analysis and interpretation of reflectivity is commonly done using the Parratt's \cite{Parrat54} or the matrix formalisms \cite{Berreman72}, assuming homogeneous optical properties throughout the constituent materials of a heterostructure.
Although, this ``slab'' approach has shown to be very successful in describing off-resonant reflectivities, it is not clear if it still holds under resonance conditions. 
This is particularly critical in single crystalline, epitaxial TMO-films and heterostructures, whose atomic structures typically realize well defined lattice planes containing the resonant scatterer.
At resonance, these atomic planes will interact very differently with the photon beam than the non-resonant regions of the material, which immediately raises the question in how far this situation can still be described using a single $n$, i.e., by assuming an optically homogeneous material. These effects are particularly important when studying any sort of electronic reconstruction at surfaces and interfaces with RXR, since they are, in fact, expected to occur on atomic length scales as well.

In this report, we investigate in detail the effects in RXR, which are caused by the rapid variations of the x-ray optical properties, which are caused by a periodic arrangement of the resonating lattice planes along the growth axis.
Using a single thin film of LaSrMnO$_4$ as a practical example, we derive analytical expressions for the reflectivity based on the Parratt's formalism in which the structure of the film is considered as layered, i.e., each plane of atoms is considered as a layer. We find that variations of $n$ at interatomic distances can have significant effects on the RXR, even in the soft x-ray range, where the wavelength of the photons is usually considered large as compared to interatomic distances. Indeed, the sensitivity of RXR to the atomic structure of a material enables to extract information about a heterostructure like interface terminations and stacking sequences, which significantly extends the capabilities of RXR. We show that from this approach, spectral information about buried interfaces can be extracted and attributed to one or several specific atomic layers.

\section{Slab versus atomic slices: theory}
\label{sec:theory}

\begin{figure}[h]
\centering
\includegraphics[width=0.47\columnwidth]{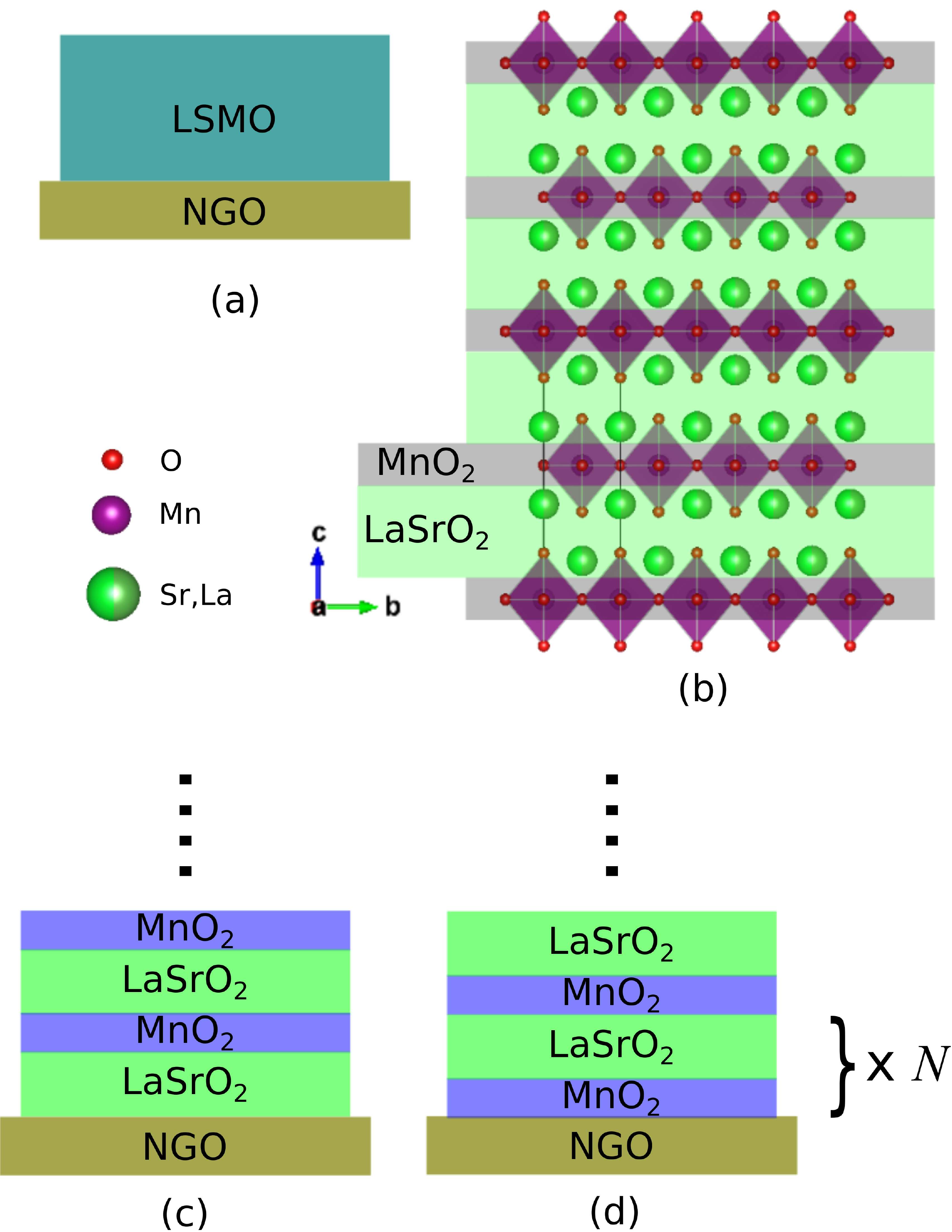}
\includegraphics[width=0.5\columnwidth]{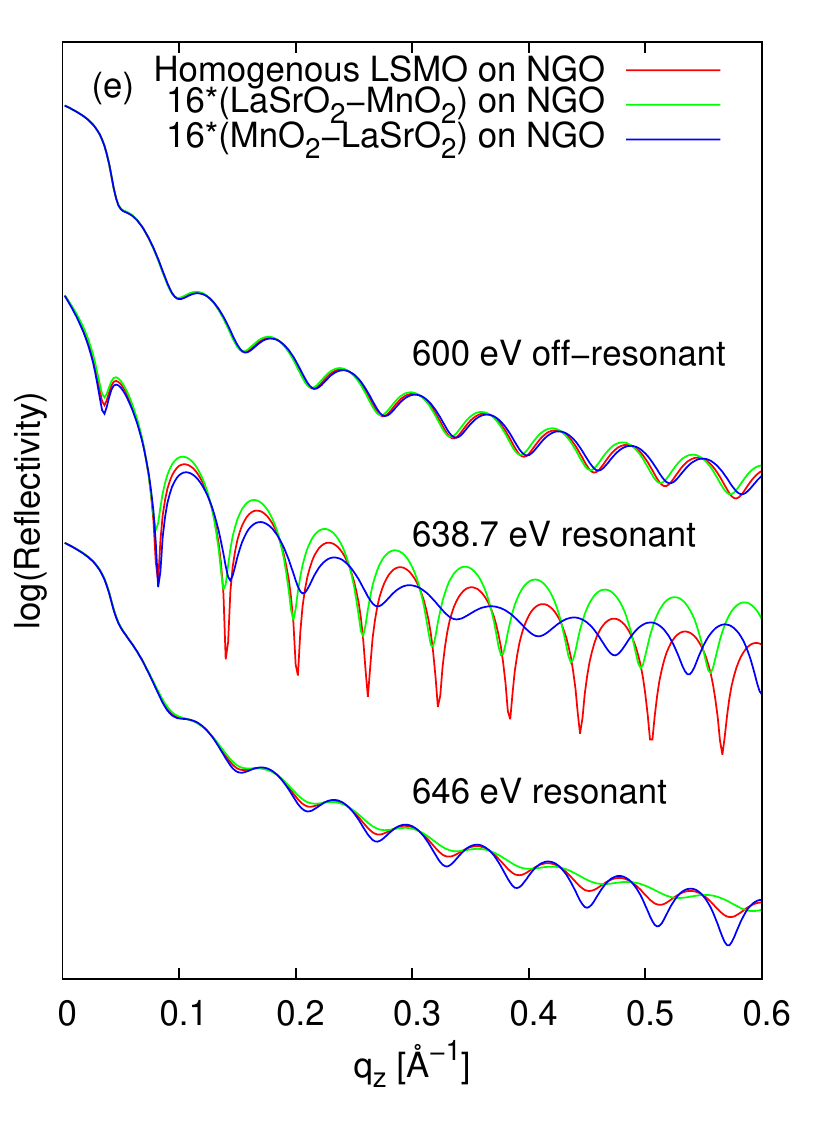}
\caption{Comparison of calculated reflectivities for an LSMO film using three different structural models. (a) homogeneous slab, (b) the LSMO crystal structure showing the different LSO and MO atomic slices. (c) and (d) LSMO film modeled as a bilayer structure for the LSO/MO and MO/LSO stacking sequences, respectively. $N$ refers to the number of bilayers, in this case we have $N=16=8$~u.c. The average density and the total thickness is the same in each case. Note that an inversion of the layer structure, i.e, LSO/MO or MO/LSO, has a dramatic impact on the reflectivity. The refractive indices of the layers used for the calculation were determined from the experimental TEY spectra of a single layer LSMO (cf. section \ref{sec:methods}).}
\label{fig:layer_comparison}
\end{figure}

When calculating the reflectivity, the crystal structure of the film and how it is simplified has important consequences for the calculated intensities, especially at resonant conditions. Before we start discussing these effects in detail, we first demonstrate how significant these effects can be using the model calculations presented in Fig. \ref{fig:layer_comparison}. Here we show the theoretical reflectivities for a 8 unit cells (u.c.) thick \lsmo~(LSMO) film grown on a NdGaO$_3$ (NGO) substrate using three different assumptions for the film structure. 

In the first model, which will be referred to as ``slab'' from now on, the conventional approach to reflectivity is applied, i.e., the LSMO film and the NGO substrate are described as slabs with homogeneous optical/electronic properties given by its refractive index $n(\omega)_{j}$ (cf. Fig.\ref{fig:layer_comparison}a). 
In the second type of model, called ``atomic slices'' in the following, additional information from the crystal structure is included. As shown in Fig. \ref{fig:layer_comparison}b, the MnO$_2$ (MO) and LaSrO$_2$ (LSO) atomic layers of LSMO are represented by considering them as thin slices with corresponding refractive indices $n(\omega)_{MO}$ and $n(\omega)_{LSO}$ and thicknesses of 1.8617~\AA{} and 4.5868~\AA{}, respectively. The thicknesses of these slices were determined using the fractional atom positions in the unit cell\cite{Ghebouli11} and the value of the experimental lattice parameter $c=12.897$~\AA{} of the film obtained from x-ray diffraction\cite{Vafaee13}.
As we will describe below, the reflectivity is much more sensitive to different stacking sequences than to the absolute value of the chosen slice thickness.
At energies close to the Mn $L$ edges, the refractive index $n(\omega)_j$ for all the layers in the slab as well as in the atomic slices approach were determined using experimental Mn scattering factors $f'$ and $f''$ determined from x-ray absorption spectra as described in the methods section (cf. Sec. \ref{sec:methods}). Also, the average LSMO density and the total film thickness (8 u.c.) is the same for each model calculation. The NGO is again described as homogeneous slab. For the atomic slices description, we further consider two different LSO and MO stacking sequences (cf. Fig.\,\ref{fig:layer_comparison}\,c,d) and compare them with the slab model in Fig.\ref{fig:layer_comparison}\,a. For the sake of simplicity and in order to focus on the differences between the models discussed here, roughness will not be considered in the following description. Instead we focus on idealized systems with perfectly sharp interfaces. The realistic case with finite interface roughness will be discussed in the analysis of our measurement data in section 4.2.

The calculated reflectivities using these three models are shown in Fig. \ref{fig:layer_comparison}\,e. As one can see, at photon energies away from the Mn L-edge, (cf. curve at 600~eV in Fig. \ref{fig:layer_comparison}e), the reflectivities are similar to one another especially at small \textbf{q$_z$} values. In strong contrast to the off-resonant region, the calculated reflectivities at resonance differ a lot depending on the model used. Furthermore, the atomic slices calculations for the different stackings  are distinctly different  as well (green and blue curves), showing that RXR is able to discriminate between different atomic stacking sequences. All reflectivity differences between the models become more pronounced at larger \textbf{q$_z$} values. 

\begin{figure}[t]
\centering
\includegraphics[width=1\columnwidth]{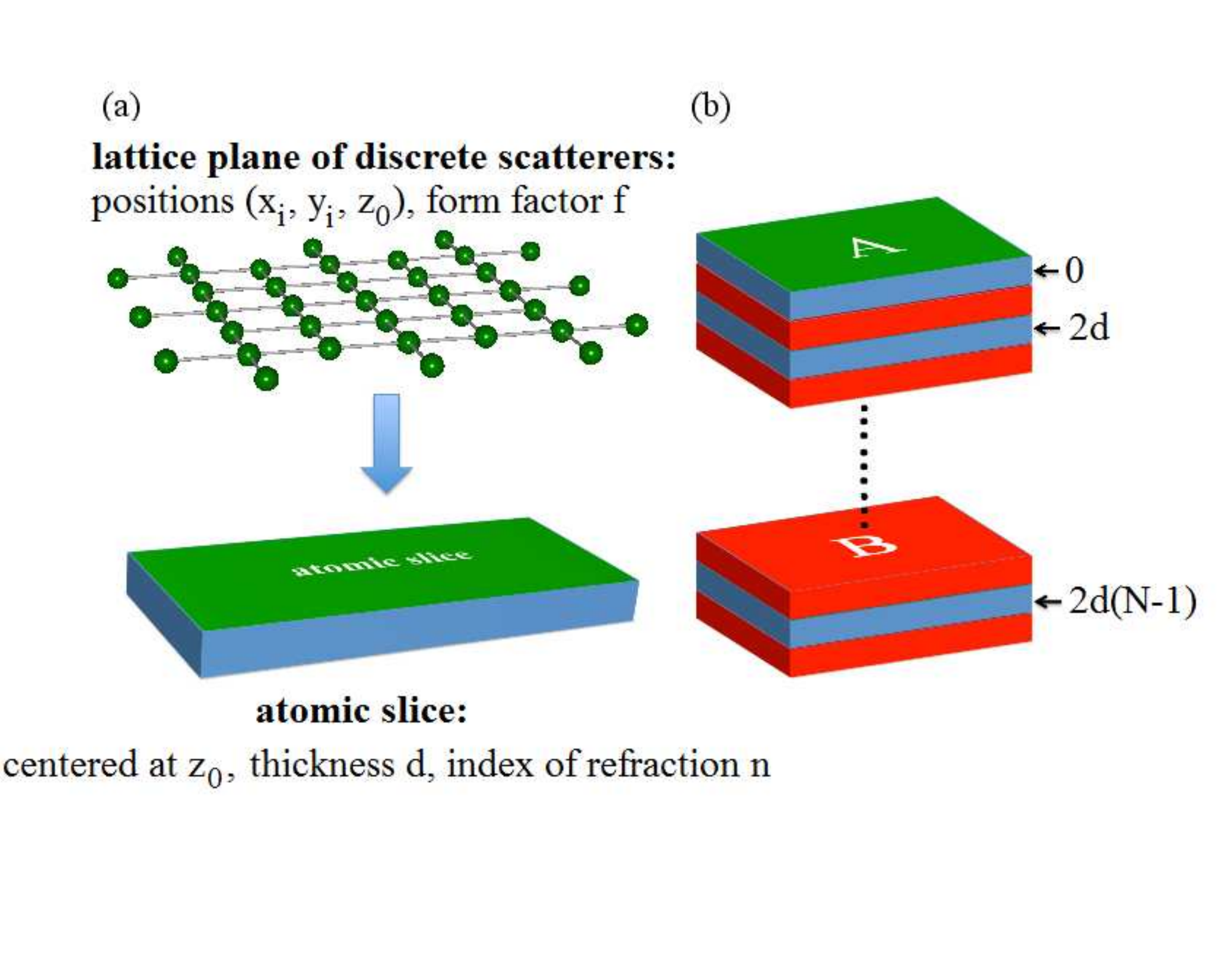}
\caption{Atomic slice model. (a) description of atomic planes of a crystal as thin homogeneous slices with finite thickness $d$. (b) material whose crystal structure consists of atomic planes of different composition A and B in the atomic slice representation.}
\label{fig:atomic-slices}
\end{figure}

At first sight the strong sensitivity of RXR in the soft x-ray range to the atomic structure of the film is surprising, because the wavelength of soft x-rays is typically considered to be large compared to the atomic structure. For this reason the atomic layers in perovskite films are usually not taken into account when analyzing soft RXR data\,\cite{Macke14,eva11,brueck11}.  
In order to better understand the results presented in Fig. \ref{fig:layer_comparison}\,e and to discuss the discrepancies observed for the above mentioned models, a closer look at the atomic slice model is required.
In this model we incorporate the atomic structure of the film into our analysis using thin homogeneous slices, as illustrated in Fig.\,\ref{fig:atomic-slices}. In this approach, the lattice planes with an area density $\eta$ of discrete sites with form factor $f$ are approximated by thin homogeneous slices with thickness $d\sim2$\,\AA, density $\rho=\eta/d$ and refractive index $n$. 

The refractive index of the latter is given by the optical theorem as $n=1+ \sum_i 2\pi \rho_i r_0 f_i/k^2$\,\cite{nielsen11}, where the sum is taken over all distinct atoms in the lattice plane. Here $r_0$ and $k$ are the classical electron radius and the x-ray vacuum wave vector, respectively.\\
For $\sigma$-polarized light with a momentum transfer vector $\mathbf{q}=(0,q_y,q_z)$ the amplitude reflectivity $r$ from a thin slice is in kinematical approximation given by\\
\begin{equation}\label{eqn:Thin_slice_complete}
r=\frac{q_z-q_{z,s}}{q_z+q_{z,s}}\left(1-{\rm exp}(iq_{z,s}d) \right)
\end{equation}
\begin{equation}q_{z,s}=\sqrt{n^2 \mathbf{q}^2-q_y^2}\approx q_z - \mathbf{q}^2/q_z \cdotp (1-n) - q_y^2 \mathbf{q}^2/q_z^3  \cdotp  (1-n)^2 + ...\end{equation}
Here $q_{z,s}$ is the z component of the momentum transfer vector in the thin slice. At this point, we will take only linear orders of ($1-n$) into account for the calculation of the reflectivity. This is justified because $(1-n) \ll1$ so that higher order terms only become relevant in $q_{z,s}$ if $q_z \ll |\mathbf{q}|$. 
Since we focus here on RXR at large  $q_{z}$, i.e., geometries far away from grazing incidence, neglecting higher orders in $(1-n)$ is justified. 
For a thin slice, we will therefore write down r as\\ 
\begin{equation}\label{eqn:1}
r=\frac{2 k^2}{q_z^2}(1-n) 
({\rm exp}(-i q_z d/2)-{\rm exp}(i q_z d/2)).
\end{equation}
The intensity reflectivity is given by $|r|^2$. 
The two phase factors in the above expression correspond to the interference of rays reflected from the top and bottom interface at $\pm d/2$ of the thin slice. This interference term has been introduced artificially by the present approximation and does not exist for a single lattice plane. It is therefore important to show that this term can be neglected, i.e., to show that the dependency of $r$ on $d$ can be neglected. To this end, we express $r$, using the above expression for $n$,
\begin{equation}\label{eqn:2}
r=-\frac{4 \pi \eta}{q_z} r_0 f  \left\{ 1-\frac{(q_z d)^2}{24}+\mathcal{O}[(q_z d)^4] \right\},
\end{equation}
which shows that the interference effects caused by the two interfaces at $\pm d/2$ do not enter, as long as $(q_z d)^2/24\ll 1$, which in turn holds as long as $d\ll\lambda/3$. This is the case for most of the soft RXR measurements where $q_z<0.5$\,\AA$^{-1}$, if $d < 5$\AA. A very similar, more general result has been obtained in Ref\,\cite{Lu2007}.

Corresponding to our example LSMO, we now consider a material with two different lattice planes separated by $d$, which are described by two different thin slices A and B, respectively (cf. Fig.\,\ref{fig:atomic-slices}\,(b)). The total amplitude reflectivity $r_{tot}$ of the whole system is the sum of the scattering from all the slices with the corresponding relative phases. Using the leading order term of Eq.\,\ref{eqn:2}, one obtains for a film with $N$ unit cells
\begin{eqnarray}\label{eqn:3}
r_{tot} & = & -\frac{4 \pi r_0}{q_z} \sum_{\nu=0}^{N-1} \left\{ \eta_A f^A + \eta_B f^B  \, e^{i q_z d}\right\}\, e^{2i q_z  d \nu}\nonumber\\
 & = & -\frac{4 \pi r_0}{q_z} \left\{ 
\frac{\overline{\eta f}}{1-e^{i q_z d}} + \frac{\delta(\eta f)}{1+e^{i q_z d} }
 \right\}  (1-e^{i q_z \Delta}),
\end{eqnarray} 
with $\Delta =2 N d$ the thickness of the film, $\overline{\eta f}= (\eta_A f^A+\eta_B f^B)/2$ describing the average scattering strength of the film material and $\delta(\eta f)= (\eta_A f^A-\eta_B f^B)/2$ representing the difference of the scattering strengths of A and B. For $q_z d \ll 1$ this can be approximated by
\begin{eqnarray}\label{eqn:4}
r_{tot} & \simeq&  -\frac{4 i \pi r_0}{q_z^2} \left\{\overline{\rho f} - \frac{i\delta(\rho f)}{2}\cdot q_z d 
\right\}  (1-e^{i q_z \Delta}),
\end{eqnarray}
where the term proportional to  $\overline{\rho f}$ corresponds exactly to what is obtained by describing the film as a single homogeneous slab without internal structure.

But from Eq.\,\ref{eqn:4} it is also clear that the latter description starts to fail as soon as $\delta(\rho f)$ is not small compared to $\overline{\rho f}$ and $q_z d \sim 1$. Indeed, in soft RXR usually one lattice plane is at resonance, while the others are not, which means that $\delta(\rho f)$ will be rather large. In addition to this, $q_{_z\,\rm max}\simeq 0.5$\AA$^{-1}$ at the transition metal $L$-edges and $d$ in transition metal oxides is typically of the order of 2\,\AA, so that at large momentum transfers $q_z d \sim 1$.  The important result of our analysis is that in the soft x-ray region, 
the atomic structure can affect the RXR signal. In other words, the corresponding phase differences can matter and, hence, the wavelength $\lambda$ can not always be considered infinitely large as compared to the lattice spacings.
Note also that at a certain resonance $\delta(\rho f)$ can be large, even for elements with similar atomic number. At resonance conditions it is in fact possible to have a finite $\delta(\rho f)$ for the same element with different electronic configurations (e.g., valence or spin state) .
According to the above discussion the contributions to $r_{tot}$, which originate from the internal atomic structure of the sample, become significant at resonance and at large momentum transfer. These two regions are exactly the most important ones when soft RXR is used for characterizing electronic reconstruction phenomena at interfaces. \\

\begin{figure}[h]
\centering
\includegraphics[width=0.3\columnwidth]{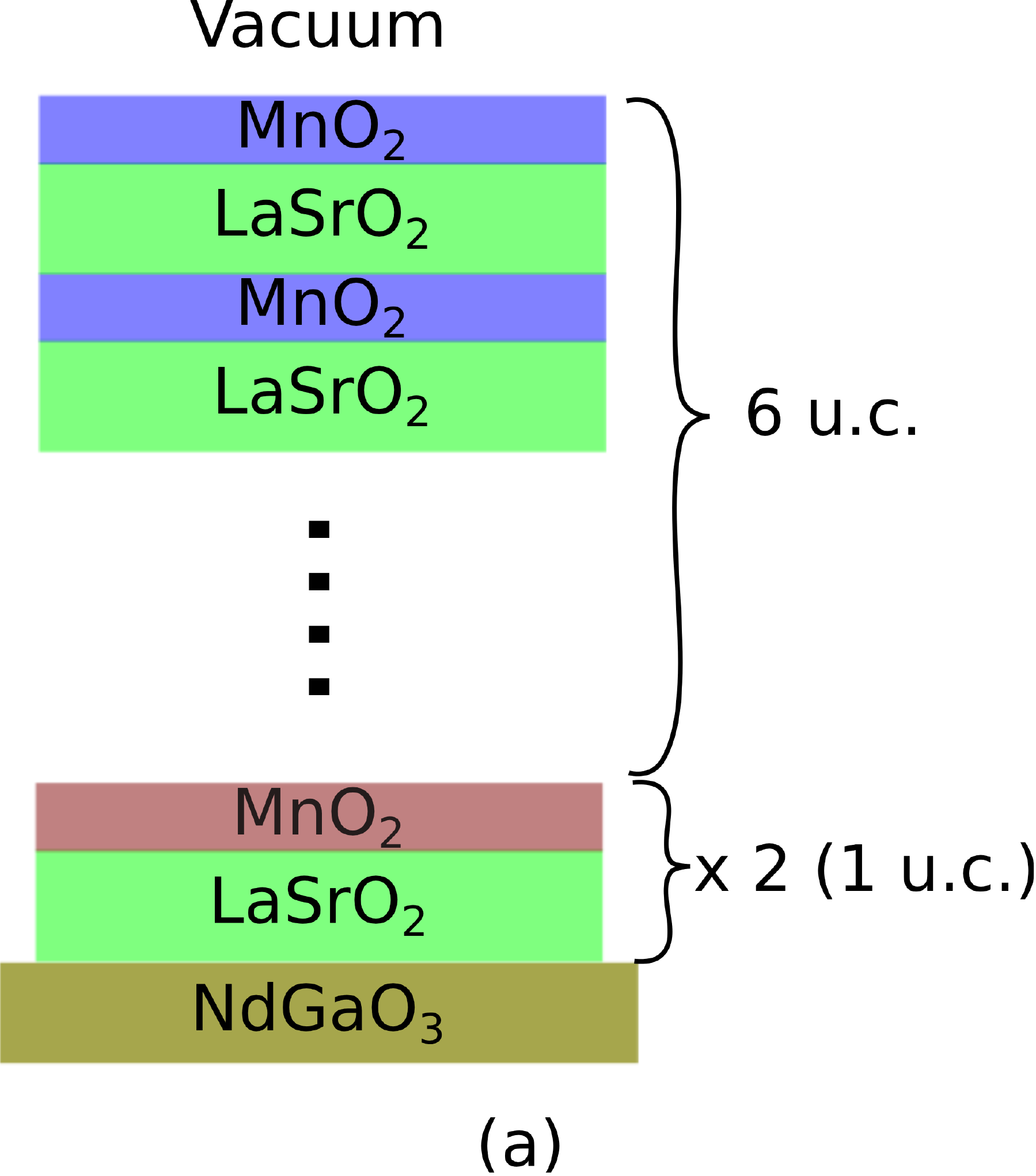}\hspace{0.5cm}
\includegraphics[width=0.57\columnwidth]{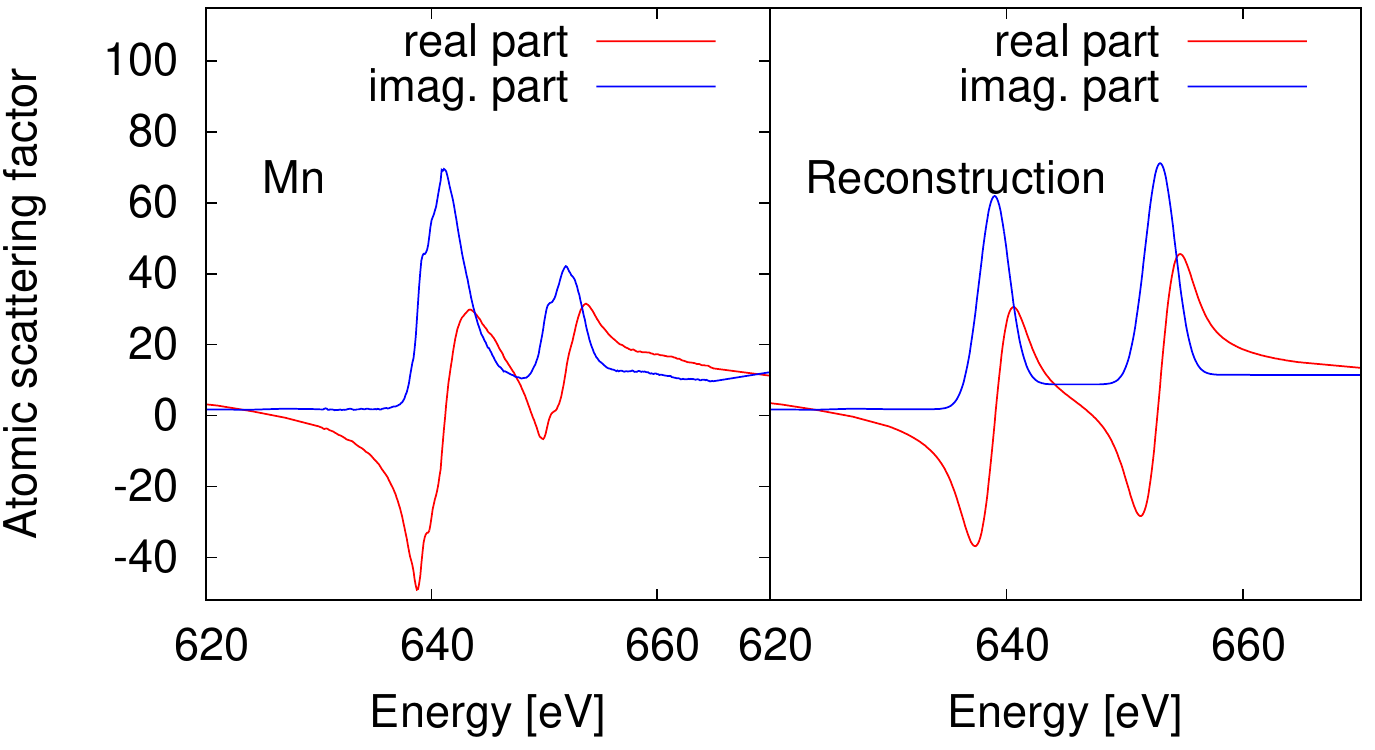}(b)
\caption{Electronic reconstruction of a 7 u.c. \lsmo~film. (a) Structural model used for the simulation of the reconstruction. For the first unit cell at the film/substrate interface Mn (shown in red) has an electronic reconstruction where 50\% of Mn is in the 3+ state and the rest is reconstructed. (b) Mn scattering factors, red and blue show the real and imaginary part of Mn$^{3+}$ and of the reconstructed Mn, which was constructed to have a Gaussian shape. The real part is obtained from a Kramer-Kronig relation.}
\label{fig:Reconstruction}
\end{figure}

\begin{figure}[h]
\centering
\includegraphics[width=1\columnwidth]{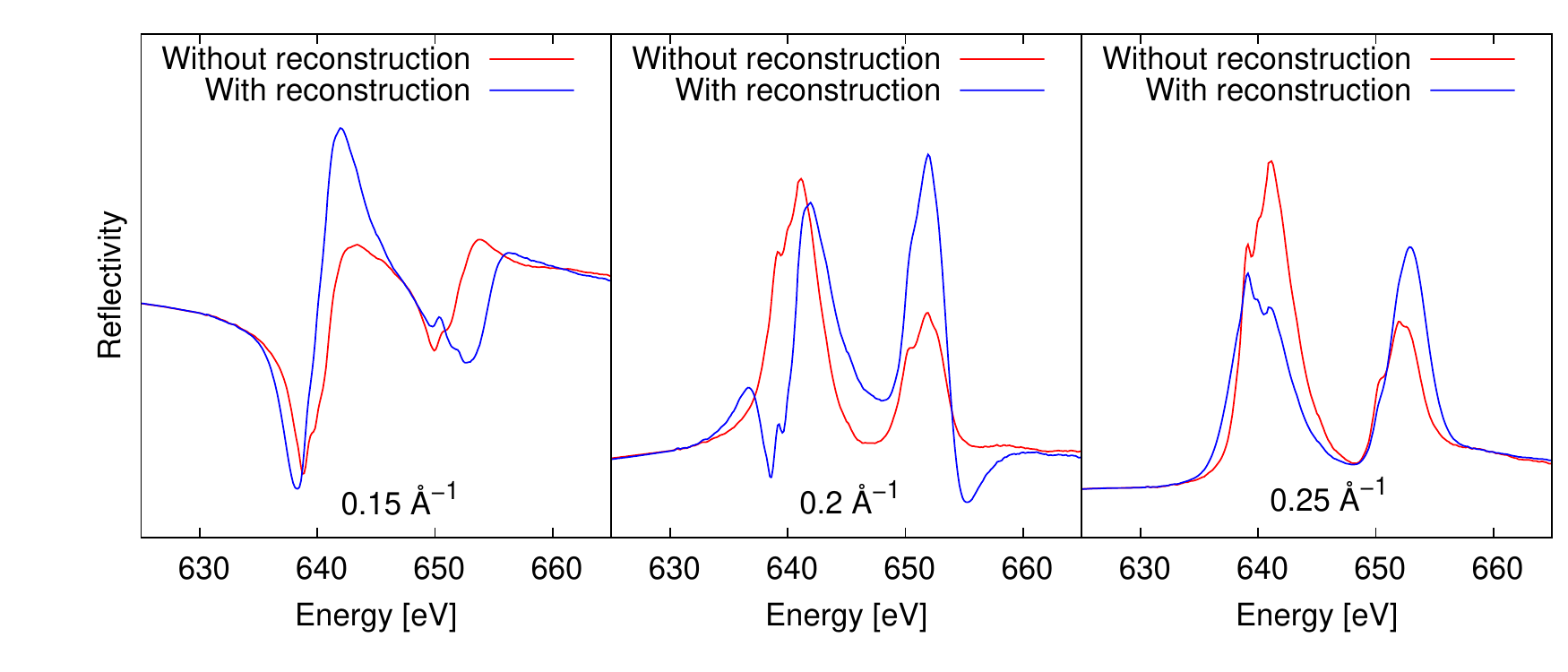}
\caption{Reflectivity as a function of energy for two selected \textbf{q}$_z$ vectors showing the effect of reconstruction.}
\label{fig:Reconstruction-sensitivity}
\end{figure}

\begin{figure}[h]
\centering
\includegraphics[width=0.8\columnwidth]{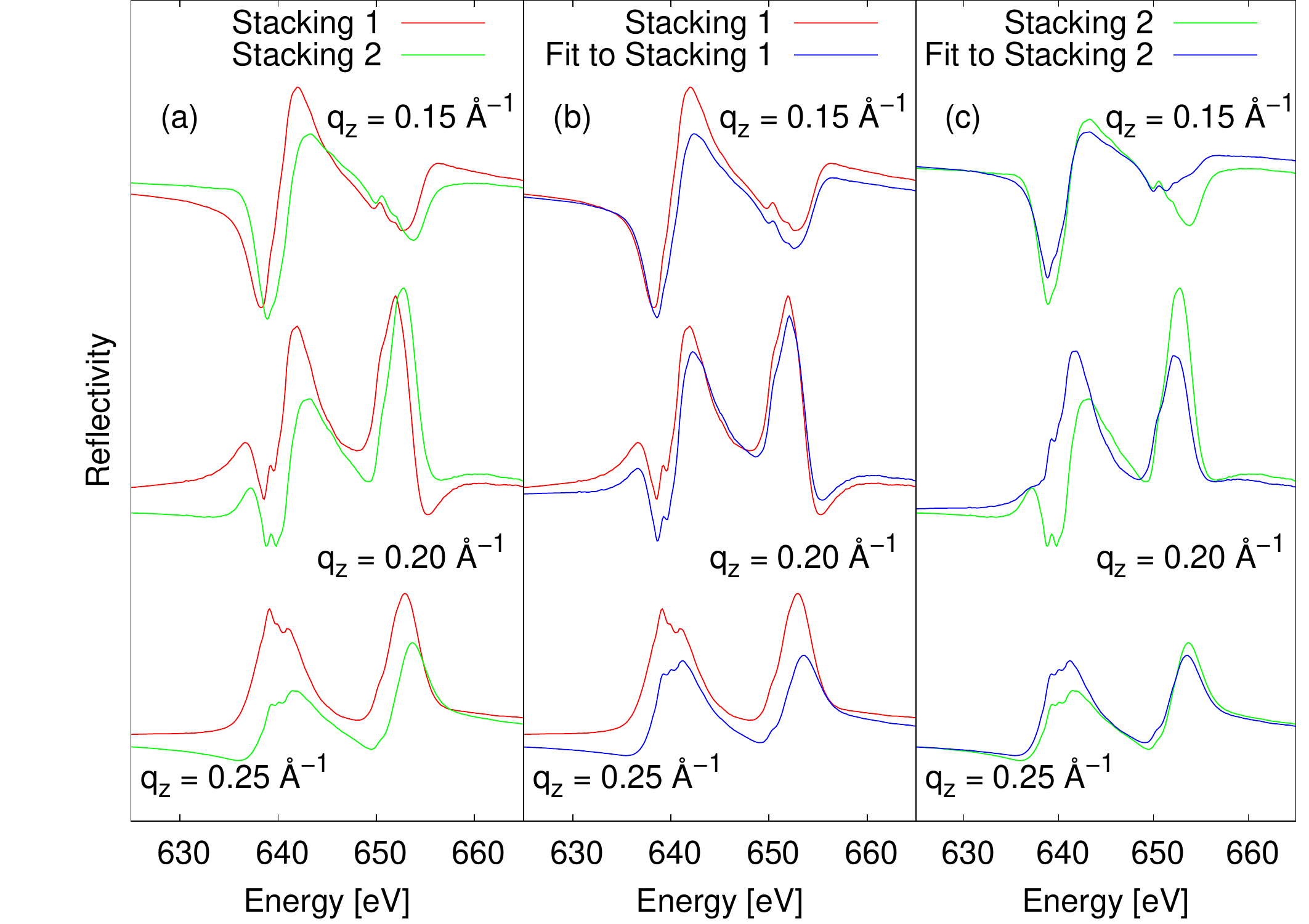}
\caption{Calculated reflectivities from an LSMO film (7 u.c thick) on an NGO substrate with an artificial electronic reconstruction at the film/substrate interface. For the sake of simplicity, the system is assumed to be perfectly crystalline with sharp interfaces. The calculated reflectivities in (a) correspond to NGO/14$\cdot$[MnO$_2$-LaSrO$_2$] (stacking 1) and NGO/14$\cdot$[LaSrO$_2$-MnO$_2$] (stacking 2). (b) and (c) show the fitting of the calculated reflectivities using a slab approach for stacking 1 and 2, respectively. For stacking 1, the slab approach yields a fairly good description of the reflectivity at low $q_z$. For stacking 2, the reflectivity lineshape can not be described very well with a slab approach. The total fitted dataset consisted of 10 reflectivities at constant energy and 10 reflectivities at constant $q_z$, for each polarization. For clarity only few reflectivities are shown.}
\label{fig:layer_comparison_fit}
\end{figure}

To asses the accuracy of the analysis of RXR data in terms of a standard slab model, we generated data sets consisting of reflectivities at energies close to the Mn $L_{2,3}$ edges for two layer stackings with the same total layer thicknesses of 7 u.c. (cf. figure \ref{fig:layer_comparison_fit}\,a). Additionally, the first LSMO unit cell on top of the NGO substrate was assumed to be reconstructed. Figure \ref{fig:Reconstruction}\,a shows the atomic slice model with the reconstruction for a film with NGO/$N \times$(LSO-MO) stacking.
For this reconstructed layer, the Mn was set to have $50 \%$ nominal Mn$^{3+}$ scattering factors (red and blue lines in figure \ref{fig:Reconstruction}\,b), as obtained from experimental XAS (cf. section \ref{sec:methods}). The other 50\% Mn was assumed to be reconstructed. The scattering factors for the reconstructed Mn are shown in figure \ref{fig:Reconstruction}\,b, where  $f''(\omega)$ was adopted to have a Gaussian line shape with an energy shift with respect to the unreconstructed case. $f'(\omega)$ is then obtained from a Kramers-Kronig relation.
As can be observed in Fig. \ref{fig:Reconstruction-sensitivity} the reconstructed layer has a strong impact on the calculated reflectivities, illustrating the high sensitivity of RXR on the atomic scale. 

The calculated reflectivities  were then fitted with a slab model. The fitting parameters were: the overall scaling factor $M$, the thicknesses of the reconstructed $\Delta_{rec}$ and LSMO $\Delta_{LSMO}$ layers and the amount $p_{rec}$ of reconstructed Mn at the interface.
The results of the fits for the two different stacking orders are shown in figure \ref{fig:layer_comparison_fit}b and c and summarized in table \ref{tab:Toyfit}.
\begin{table}[h]
\centering
\begin{tabular}{l|c|c|c}
\hline\hline
Parameters			& Start values	& Fit to Stacking 1	& Fit to Stacking 2 \\
\hline\
$M$				&1		&0.81			&1.24 \\
$\Delta_{LSMO}$\hfill[\AA{}]	&77.38		&79.76			&75.18 \\
$\Delta_{rec}$\hfill [\AA{}]	&12.897		&11.53			&13.83 \\
$p_{rec}$\hfill [\AA{}]		&0.5		&0.45$\pm$0.13		&0.54$\pm$0.22 \\
$\chi^2$			& -		&0.59			&1.40 \\
\hline\hline
\end{tabular}
\caption{Results of the fits of the calculated reflectivities for two different bilayer stacking of LSMO using the slab approach.}
\label{tab:Toyfit}
\end{table}
As seen in the table, the resulting thicknesses of the different layers are close to the starting values within an error of $\sim$1-2~\AA{}.
Also, the information regarding the amount of Mn reconstruction is quite close to the original value, however, the relative errors for the fitted values of 30-40\% are quite significant.

Based on this analysis, two main conclusions can be drawn. First, fitting RXR data using the slab approach can yield a fairly good description of the experiment. Notwithstanding, the use of the slab model introduces errors, since it simplifies the real material and its reflectivity by overlooking the contribution of the internal structure of the film. This can have important consequences: as can be seen in table\,\ref{tab:Toyfit}, the fit to stacking 1 yields a reasonable $\chi^2$ value, indicating a good fit. In contrast to this, the resulting $\chi^2$ for the fit to stacking 2 is about 2.5 times larger than that of stacking 1, implying dubious fit results and parameters. The reflectivity curves, calculated in the atomic slices approach for stacking 2, therefore cannot be described well by a slab model, which neglects the internal structure of the material.
Second, 
by taking into account information about the lattice structure and setting up a corresponding atomic slice model enables to retrieve important information, such as the stacking sequence,
which is lost when the slab approach is implemented. Therefore, a more accurate description of the experiment is obtained when utilizing the atomic slices model. 

\section{Methods}
\label{sec:methods}

A (001) oriented LaSrMnO$_4$ (LSMO) film was grown epitaxially on a (110) oriented NdGaO$_3$ substrate by the pulsed laser deposition technique. Details on the sample preparation and characterization can be found elsewhere \cite{Vafaee13}.

\begin{figure}
\centering
\includegraphics[width=0.9\columnwidth]{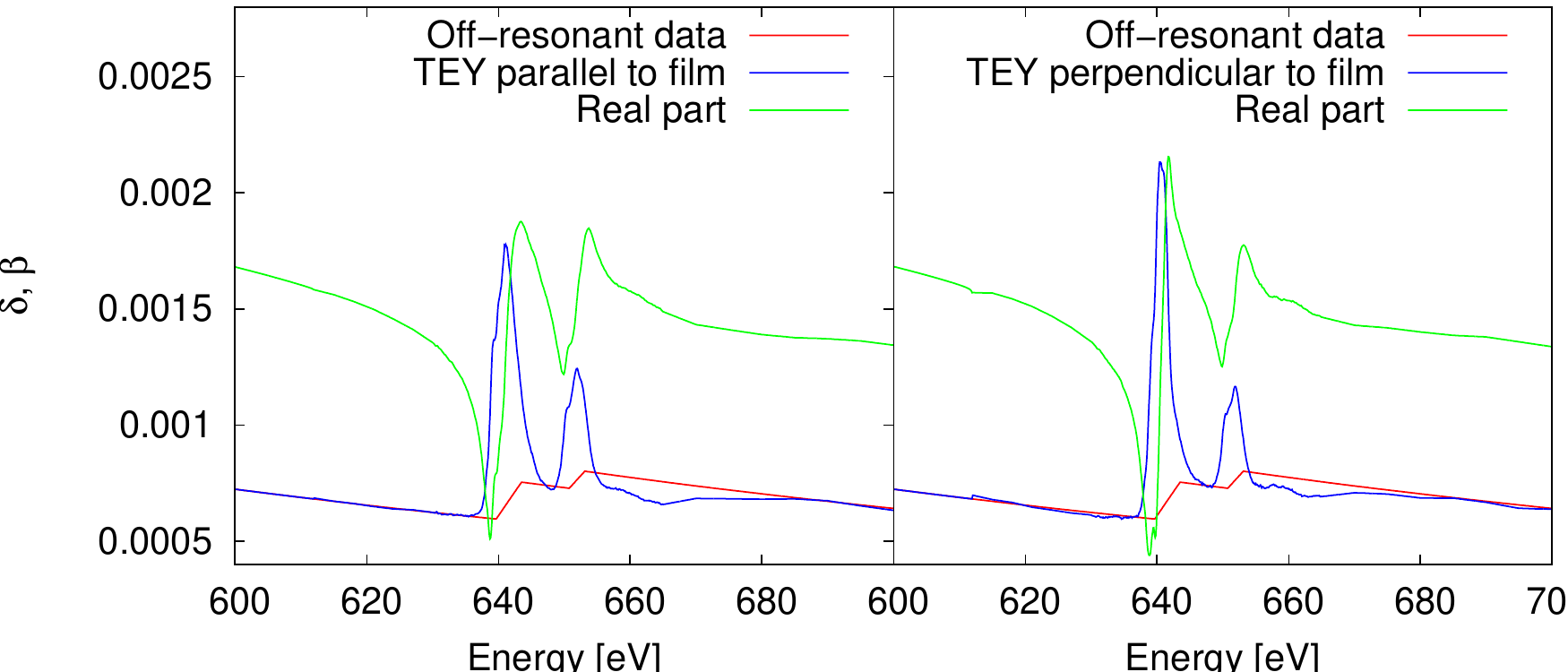}
\caption{The optical constants $\delta$, $\beta$ of LSMO at the Mn L$_\mathsf{2,3}$ edge. Determination was done as described in the work by Macke \textit{et al.} \cite{Macke14}.}
\label{fig:FitTEY2}
\end{figure}

X-ray absorption spectroscopy (XAS) and resonant x-ray reflectometry experiments have been performed at the 10ID-2 (REIXS) beamline of the Canadian Light Source (Saskatoon, Canada) \cite{hawthorn11}, using linearly $\sigma$- and $\pi$-polarized light. 
The XAS measurements were carried out in the total electron yield (TEY) mode at two different scattering geometries and incoming beam polarizations in order to extract absorption spectra corresponding to the directions in-plane and out-of-plane, i.e., $E\parallel c$ and $E\perp c$ of the LSMO, respectively. The absorption was measured at the O $K$, Mn $L_{2,3}$ and La $M_{4,5}$ absorption edges. 
The RXR experiments were carried out at energies around the Mn $L_{2,3}$ and La $M_{4,5}$ absorption edges. The reflected intensities were collected in the fixed energy (fixE) and the fixed \textbf{q}$_z$ (fixQ) modes. The fixE consist of \textbf{q}$_z$-scans ($\theta-2\theta$) carried out at a fixed photon energy, whereas the fixQ refers to energy scans at a fixed scattering vector \textbf{q}$_z$. The selected \textbf{q}$_z$ vectors correspond to maxima and minima of the thickness oscillations taken from the fixE reflectivity curve measured at 641~eV. 
Determination of the imaginary part of the Mn scattering tensor and the scalar atomic scattering factors for La, Nd, Ga, Sr and O was done as described in the work by Macke \textit{et al.} \cite{Macke14}. This is, the parallel and perpendicular components of the absorption spectra obtained from the XAS measurements were scaled to non-resonant tabulated values \cite{chantler71}. 
The X-ray absorption was measured in between the Mn $L_{2,3}$ edge and the La $M_{4,5}$ edge to avoid any influence of near-edge oscillations on the scaling.
The real part is then obtained by performing a Kramers-Kronig transformation. 
Since all experiments were performed at room temperature, i.e, well above the N\'{e}el temperature (in bulk $T_N=127K$\cite{Baumann2003}), there is no long-range magnetic ordering of the Mn-moments. In this case, the scattering matrix can be assumed to be diagonal\,\cite{haverkort-PRB2010}. From the atomic scattering factors we could calculate the dielectric tensor $\boldsymbol\epsilon$ and the refractive index $n(\omega)=1-\delta(\omega)+i\beta(\omega)$ of the film as shown in figure \ref{fig:FitTEY2}. The specular reflectivity was calculated with the Parratt's \cite{Parrat54} and matrix \cite{Berreman72} formalisms using the ReMagX suite \cite{Macke2014}. 

X-ray photoelectron spectra (XPS) were recorded using PHI VersaProbe 5000 spectrometer with monochromatic Al\,$K\alpha$ ($h\nu$\,=\,1486.6\,eV) radiation at pass energy of 23.5\,eV. 
In order to characterize the chemical composition of the film surface, spectra were collected as a function of the take off angle between the surface of the sample and the axis of photoelectron detector. Binding energies of the spectra were calibrated with an adventitious carbon C\,1$s$ emission line at 284.8\,eV. The XPS spectra were analyzed using XPSPEAK 4.1 software after background subtraction by the Tougaard method. The shape of the characteristic peaks in all spectra was considered symmetric with a combination of 30\% Lorentzian-Gaussian profile.

\section{Experimental results and discussion}
\label{sec:results}

In order to determine a realistic model for the studied heterostructure, we characterized the film surface experimentally using XPS and determined a realistic parametrization for the  LSMO/NGO interface.

\subsection{Surface characterization and description of substrate/film interface}
\label{sec:addlayers}

\begin{figure}[h]
\centering
\includegraphics[width=0.57\columnwidth, height=220px]{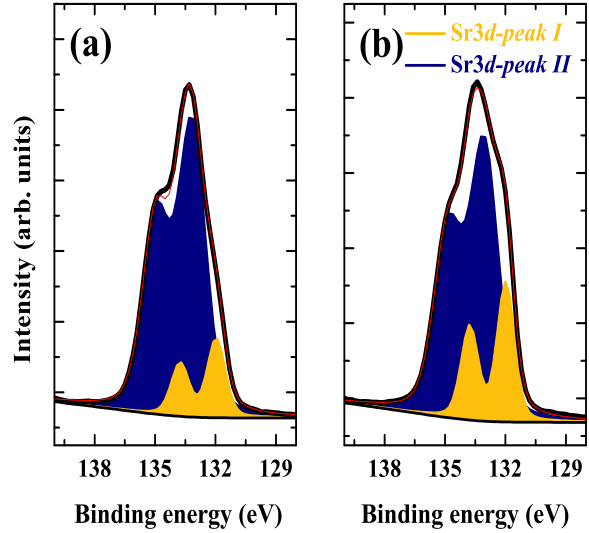}
\caption{Angle-resolved XPS spectra of Sr-3d for LaSrMnO$_4$ thin film deposited on NGO (1 1 0) substrate measured at the angle of (a) 20\,$^{\circ}$ and (b) 45\,$^{\circ}$. }
\label{fig:XPS}
\end{figure}

The existence of a SrO layer on the sample surface is revealed by the XPS data shown in \ref{fig:XPS}a and \ref{fig:XPS}b, where the take off angle dependence for Sr-3$d$ photoemission lines is presented.
As it can be seen, Sr-3$d$ shows two sets of doublet peaks shifted by $\sim 1.1$~eV in binding energy indicating two different Sr-O bonds. $Peak~II$ shown in blue has been attributed to SrO and $peak~I$ (yellow) originates from Sr-O bond in the LSMO structure \cite{Bertacco02}. 
Since the $peak~II$ contribution to the spectrum measured at 20\,$^{\circ}$ is larger in comparison with the one measured at 45\,$^{\circ}$, and considering that XPS measurements at small angles are more surface sensitive, it can be concluded that SrO segregates at the surface. Such a SrO segregation layer is commonly found on the surface of manganites \cite{Fister08,Katsiev09,Abellan11,Li12,Poggini2014}. We therefore included a SrO surface layer into the model for the reflectivity.

In addition to this segregation layer, a surface adsorption layer on top of the SrO was considered based on the fact that the sample has been exposed to air and contaminants such as water molecules, carbon etc., can be adsorbed on the sample surface. As a simplification, we consider only the scattering from oxygen in the adsorption layer. 

\begin{figure}
 \centering
\includegraphics[width=0.5\columnwidth]{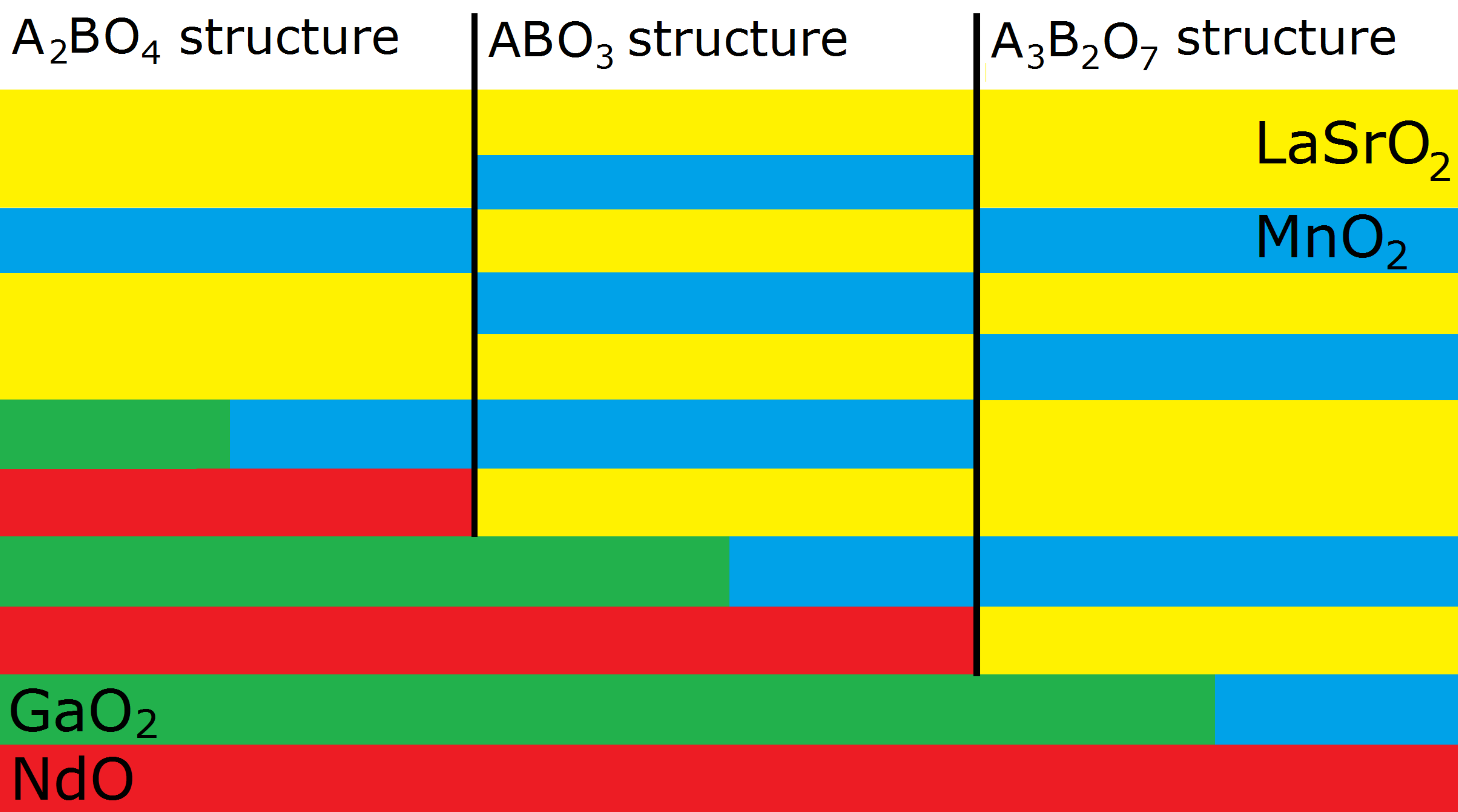}
\caption{Antiphase boundaries of the LSMO structure when growing on an NGO substrate. Once a given thickness is reached, the film grows epitaxially.}
\label{fig:intermixing}
\end{figure}

Furthermore, it is known from the growth of Ruddlesden-Popper compounds on a substrate that if the c-axis 
of the thin film is larger than that of the substrate, the substrate terraces will lead to antiphase boundaries in the film \cite{Haage97,Malik12}. Such a scenario is depicted in figure \ref{fig:intermixing}. From the figure it can be seen that it is relatively easy to heal antiphase boundaries in LSMO by inserting related compounds like LaMnO$_3$ and La$_{1.5}$Sr$_{1.5}$Mn$_2$O$_7$ into the stack. Once the LSMO thickness of the film increases some domains become dominating and finally the majority LSMO domain grows epitaxially. The presence of a minority phase containing single atomic layers of LaSrO$_2$ in this system was also demonstrated by means of x-ray diffraction. \cite{Vafaee13}
To model such an interface, a \textit{transition} layer was included with a refractive index that was considered to be a linear combination of that of NGO and LSMO. For this layer we have defined $n_{int}=(1-p)\cdot n_{NGO}+p\cdot n_{LSMO}$, where the factor $p$ was fitted during the analysis of the reflectivities. 

\subsection{Fit of the reflectivity data}
\begin{figure}
 \centering
\includegraphics[width=\columnwidth]{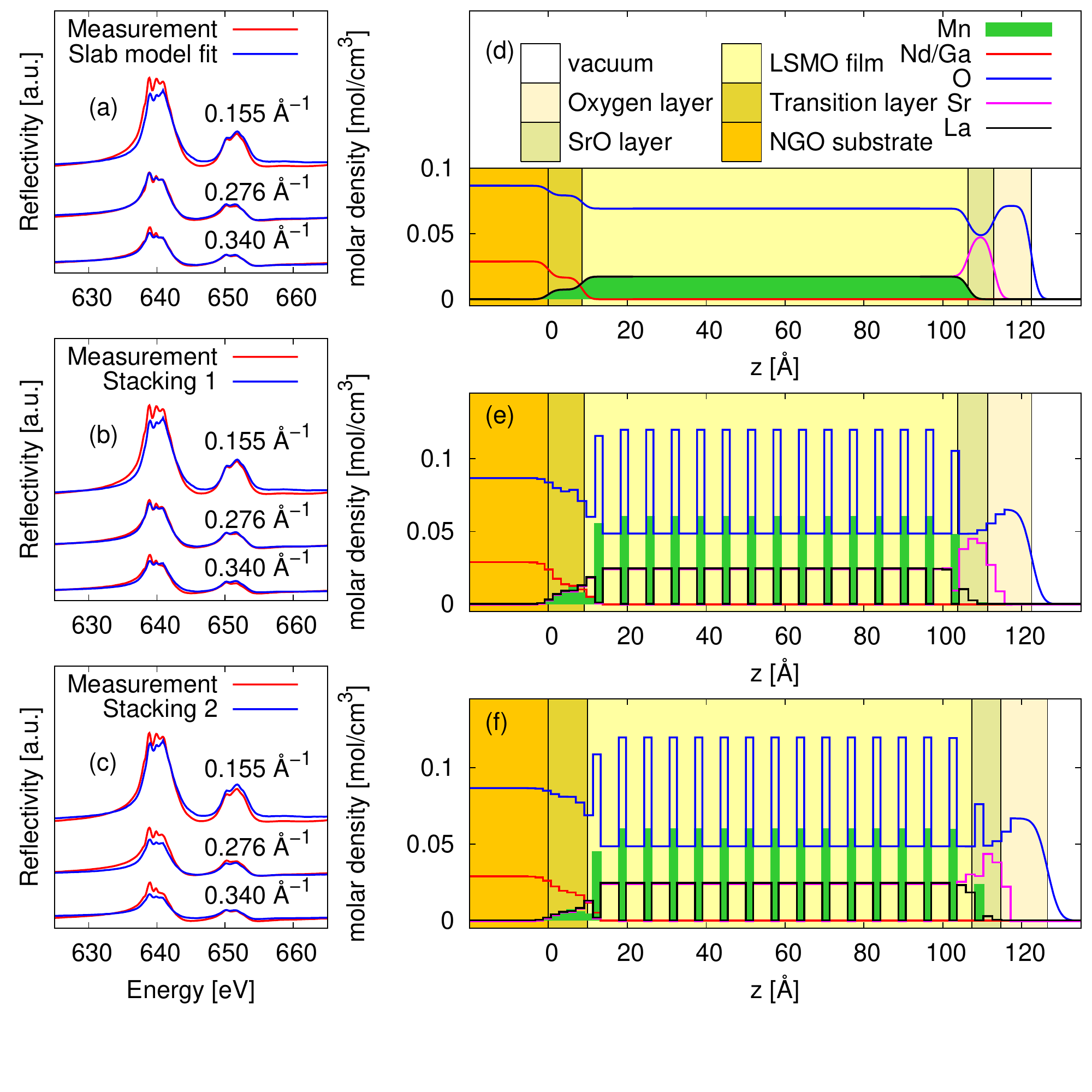}
\caption{Selected resonant x-ray reflectivities measured close to the Mn $L_{2,3}$ edges. (a) shows the fit result based on the slab approach, (b) shows a fit result based on an atomic slices approach with the lowest $\chi^2$ stacking, whereas (c) shows a fit result where the LSMO/SrO termination was fixed to be at the LaSrO$_2$ layer. (d), (e), and (f) show the elementary density profile throughout the film thickness for the corresponding models. The lines showing the interface between the \textit{transition}/LSMO and LSMO/SrO layers corresponds to the $z_0$ value of the error function (Eq. \ref{eq:err-func}) at that interface.}
\label{fig:comparison_exp}
\end{figure}

The model used for the analysis of the experimental reflectivities consists of an LSMO thin film on an NGO substrate with an NGO/LSMO \textit{transition} layer, a top SrO segregation layer, which itself is covered by a contamination layer (cf. section \ref{sec:addlayers}).
Figure \ref{fig:comparison_exp} shows representative experimental reflectivities (red) together with fits (blue) corresponding to the slab (Fig. \ref{fig:comparison_exp}a) and atomic slices approach (Fig. \ref{fig:comparison_exp}b and c). The fit results are summarized in table \ref{tab:fit-res}.
During the fitting procedure, all the measured reflectivities at both $\sigma$ and $\pi$ polarization were fitted simultaneously in order to get a self-consistent result. The fitting parameters were the overall scaling factor $M$, the background intensity as a fraction of the total beam intensity, interface positions $z_{0,j}$ of all layers, the adsorbate density $\rho_O$ at the contamination layer and the ``intermixing'' parameter $p$ of the interface layer (cf. section \ref{sec:addlayers}). For the slab approach, all the layer thicknesses were fit parameters and a Nevot \& Croce roughness approximation \cite{nevot80} was used at all interfaces.

\begin{figure}
\centering
\includegraphics{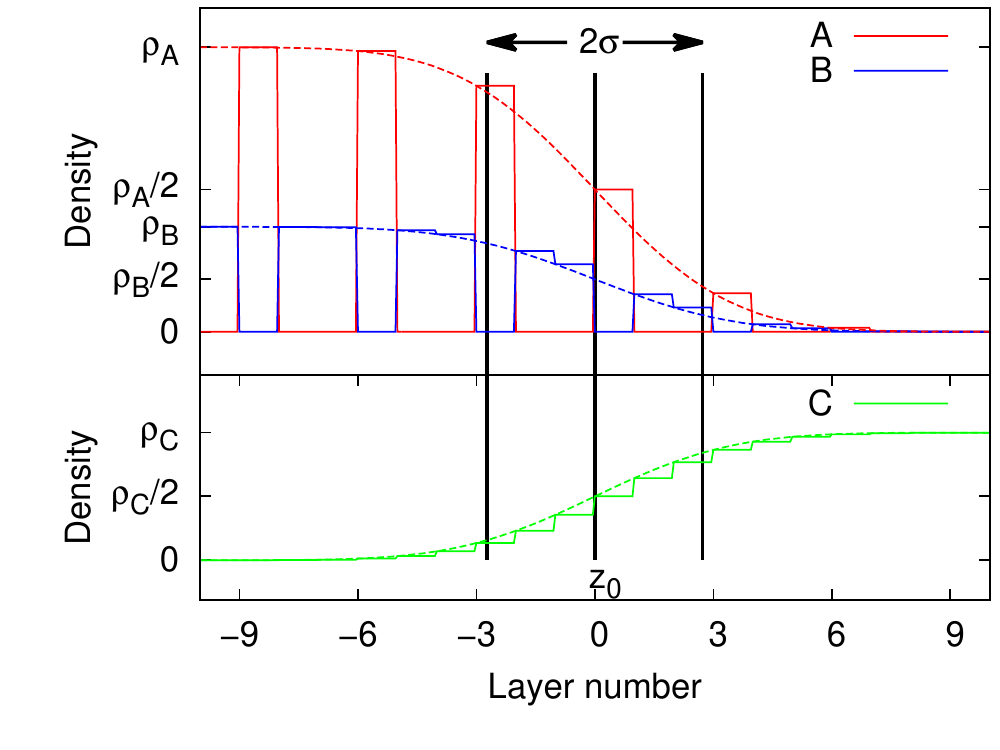}
\caption{Parametrization of a generic interface according to Eq. \ref{eq:roughness_param}. Rough interfaces of layered compounds can be modeled with the two continuous parameters roughness $\sigma$ and termination point $z_0$.}
\label{fig:roughness}
\end{figure}

In case of the the atomic slices approach, the Nevot \& Croce roughness approximation can  no longer be applied. Instead we use a new approach that allows for a smooth transition between different materials, preserves crystallinity and converges to Nevot \& Croce roughness in the limit of high roughness.
This approach is sketched in Fig \ref{fig:roughness}. Lets assume we have an interface of two layered materials. Material L consists of two layers with composition A and B whereas material S consists of layers with composition C. The transition from 100\% L into 100\% S is parametrized using an envelope error function $f(z)={\rm erf}(z,\sigma)= (\sigma \sqrt{2 \pi}) \int_{-\infty}^z exp(-\zeta^2/2\sigma^2) d\zeta$. Here $\sigma$ defines the width of the transition, i.e., the roughness. We now define the relative abundance of material S in a certain position $z$ through the interface as
\begin{equation}
S =  1- {\rm erf}\left((z-z_0)\,d_{\rm av},\sigma\right)
\label{eq:err-func}
\end{equation}
where $z$ is an integer atomic slice number. In our approach the interface location is given by $z_0$, where the relative abundance of material L and S is 50\%, respectively. $d_{av}$ is the average atomic thickness between material L and S.
Correspondingly, the relative abundance of L is defined as $L=1-S$. Note that this parametrization allows us to choose a given film termination at any interface by fixing the $z_0$ value.
To illustrated the interface definition, let us consider the LSMO/SrO interface as an example. The SrO is segmented into slices with density $\rho_{\rm SrO}=5.01~g/cm^{3}$ and thickness $d_{\rm SrO} =2.31$~\AA{}. The relative abundance of SrO (S) at a certain position through the interface is calculated using Eq. \ref{eq:err-func}. This allows us to determine the abundance of LSMO (L).

Once $L$ and $S$ are determined one can then calculate new densities and thicknesses, i.e., optical constants, for the layers at the interface by summing the contributions of the SrO and LSMO layers as follows
\begin{equation}
\rho_{\rm Mn}(z)=L \cdotp \rho_{\rm Mn, \;MnO_2-Layer} \; \; \; {\rm if} \;  z/3 \in \mathbb{Z},\; {\rm else} \;\rho_{Mn}( z) = 0
\label{eq:roughness_param}
\end{equation}

\begin{equation}
d(z)=L \cdotp d_{\rm MnO_2} + S \cdotp d_{\rm SrO}  \; \; \; {\rm if} \;  z/3 \in \mathbb{Z}
\end{equation}
The definitions for the other atom densities and atom slice thicknesses are set up in the same way.
During the fitting procedure, the interface positions $z_{0,j}$ for all interfaces $j$ can then be fitted thus yielding the interface termination. Correspondingly, the film thickness is determined from the relative distance between $z_{0,j}$ of the top and bottom interfaces.

Like the Nevot \& Croce roughness approximation, this method should only be applied if the roughness is much smaller than the film thickness, i.e., in the case of sharp and well-defined interfaces. We found that this requirement is fulfilled for our system.
In order to obtain stable fit results, we had to assume the same roughness $\sigma$  for all interfaces. Although this obviously needs not to be fulfilled in the real heterostructure, the corresponding model fits well to the experimental results (see below), indicating that the roughnesses of the different interfaces in the real material are indeed similar. 

\begin{figure}
 \centering
\includegraphics[width=0.7\columnwidth]{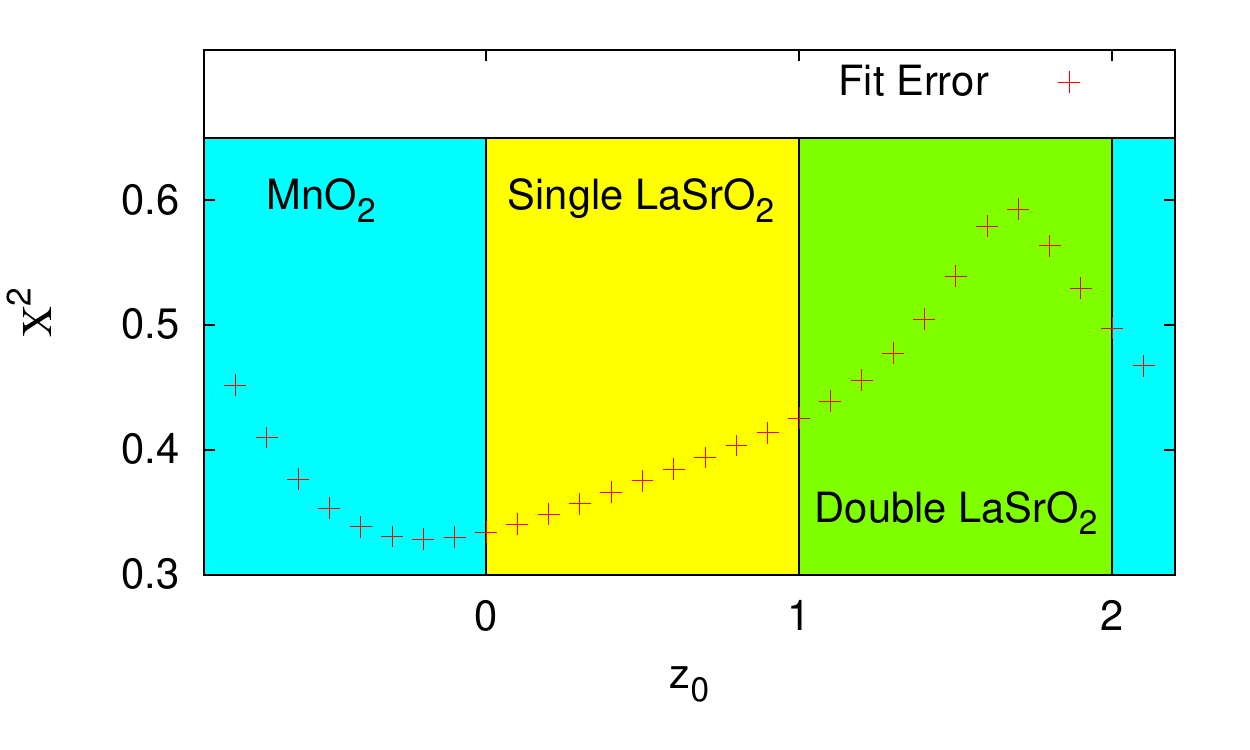}
\caption{Fit error $\chi^2$ for different LSMO terminations $z_0$ at the LSMO/SrO interface. The lowest $\chi^2$ is obtained for a MnO$_2$ terminated film.}
\label{fig:terminations_fit}
\end{figure}

Figure \ref{fig:terminations_fit} shows the resulting $\chi^2$ for different terminations ($z_0$) of the LSMO at the LSMO/SrO interface. As shown in the figure, the lowest $\chi^2$ is obtained when the LSMO  at the interface with SrO is terminated with an MnO$_2$ atomic layer. In comparison, the highest  $\chi^2$, i.e., the poorest fit to the data, is obtained when the termination is a  LaSrO$_2$ double layer. The resulting fits for these two extreme cases are shown in Fig. \ref{fig:comparison_exp}b and c, respectively and listed in table \ref{tab:fit-res}.
This analysis allows to conclude that the best fit to the experiment is given by the case in which the LSMO film is MnO$_2$ terminated. 
Regarding the interface with the substrate, although the NGO/LSMO termination is less sharp there are two main conclusions we get from the fit results. 
First, the thickness of the \textit{transition} layer is $\sim$9-10~\AA{} (cf. table \ref{tab:fit-res}), implying  that the region where antiphase boundaries appear is less than one LSMO unit cell in length. This, together with the obtained small roughness value ($\sim$2~\AA{}), shows that the NGO/LSMO-interface is very sharp . Second, the fits yield that the first layer that grows with few antiphase boundaries on the NGO is a single LSO layer. This is better seen in the elementary density profiles in Fig. \ref{fig:comparison_exp}e. The lines showing the interface between the \textit{transition}/LSMO and LSMO/SrO layers corresponds to the $z_0$ value of the error function at that interface. 

\begin{table}
\centering
\begin{tabular}{l|c|c|c}
\hline\hline
Parameters 				& Slab 			&Atomic slices 		&Atomic slices\\
					& 			&Stacking 1 		&Stacking 2\\
\hline
$M$ 					&2.16 $\pm$ 0.05 	&2.08 $\pm$ 0.05 	&1.759 $\pm$ 0.032\\
$\Delta_{\rm trans}$\hfill [\AA{}]	&8.51 $\pm$ 0.13 	&9.10 $\pm$ 0.25 	&9.90 $\pm$ 0.21\\
$\Delta_{\rm LSMO}$\hfill[\AA{}] 	&97.92 $\pm$ 0.24 	&94.64$\pm$ 0.24 	&97.50 $\pm$ 0.10\\
$\Delta_{\rm SrO}$\hfill [\AA{}] 	&6.48 $\pm$ 0.11 	&7.67 $\pm$ 0.17 	& 7.29 $\pm$ 0.27 \\
$\Delta_{\rm O}$\hfill [\AA{}] 		&9.46 $\pm$ 0.14 	&11.09 $\pm$ 0.27 	&11.83 $\pm$ 0.34\\
$\rho_{\rm O}$\hfill [$g/cm^3$] 	&1.142 $\pm$ 0.025 	&1.040 $\pm$0.024	&1.071 $\pm$ 0.021 \\
$\sigma$\hfill [\AA{}] 			&1.47 $\pm$ 0.12	&2.11 $\pm$ 0.08 	&2.36 $\pm$ 0.14 \\
$p$\hfill 				&0.427 $\pm$ 0.013 	&0.548 $\pm$ 0.013	&0.359 $\pm$ 0.013 \\
background/10$^{-7}$ 			& 3.12$\pm$ 0.13 	& 2.93 $\pm$ 0.13	& 2.65 $\pm$ 0.11 \\
$z_0$ (LSMO/SrO) 			&- 			& 2.81 $\pm$ 0.05	&1.7 (fixed) \\
$\chi^2$ 				&0.35 			&0.33 			&0.59\\
\hline\hline
\end{tabular}
\caption{Fit results of experimental reflectivity data with different structural models. Note that a model that neglects atomic slices completely is still better than a model with wrong stacking. The given errors were taken from the fit covariance matrix.}
\label{tab:fit-res}
\end{table}

Comparing the resulting fits using both slab and atomic slices (Figs. \ref{fig:comparison_exp} a,d and b,e, respectively) it is not obvious which yields the best description of the data. From a qualitative point of view, both models reproduce most of the features such as the thickness oscillations. Also their fit errors $\chi^2$ are similar as shown in table \ref{tab:fit-res}. 
Still, the fit using atomic layers has important information that is completely lost when the slab approach is used. This is the stacking sequence of the LSO and MO bilayers. 
Such information is very import not only for understanding the film growth but also in order to explain physical phenomena, which are determined by terminations, such as the effects at the LAO/STO interface. This can only be retrieved by describing the film in terms atomic layers that form the crystal structure. 

The power of this approach is better seen by comparing the fits using the atomic slices approach with two different stacking sequence of the bilayers, i.e., figures \ref{fig:comparison_exp}b and c. Their $\chi^2$ values differ considerably, thus showing that only stacking 1 yields a proper description of the experiment.
So far, our model assumes a single Mn species. However, as discussed above, our analysis already shows that there is a transition layer between the NGO-substrate and the LSMO-film as well as a SrO-layer at the top of the LSMO-film. The Mn-sites in these regions are therefore located in a different chemical environment than the ones inside the film. The properties of the Mn valence shell can hence be expected to be different in these different regions of the heterostructure. In addition to this, also symmetries are broken at interfaces and polar structures like LSMO may exhibit a so-called electronic reconstruction.

To obtain depth-resolved spectroscopic information about the Mn in the various layers, we followed a Kramers-Kronig constrained variational fit approach that was outlined in great detail by Kuzmenko \cite{Kuzmenko2005} and subsequently applied to the case of resonant X-ray reflectivity \cite{Stone2012}. In this approach, local changes of the complex dielectric function $\epsilon$ are modeled in the imaginary part by a mesh of triangular functions with tunable height. The real part of  $\epsilon$ can then be easily determined by adding the known real parts of the fitted triangular functions. Choosing every second data point in energy as a mesh, we were able to simultaneously fit the in-plane ($\epsilon_{xx}=\epsilon_{yy}$) and out-of-plane ($\epsilon_{zz}$) component of the dielectric function employing a Gauss-Newton type fit algorithm.The computing time needed for the fitting is short and convergence typically occurred within a few minutes on a modern desktop computer.

\begin{figure}
 \centering
\includegraphics[width=1\columnwidth]{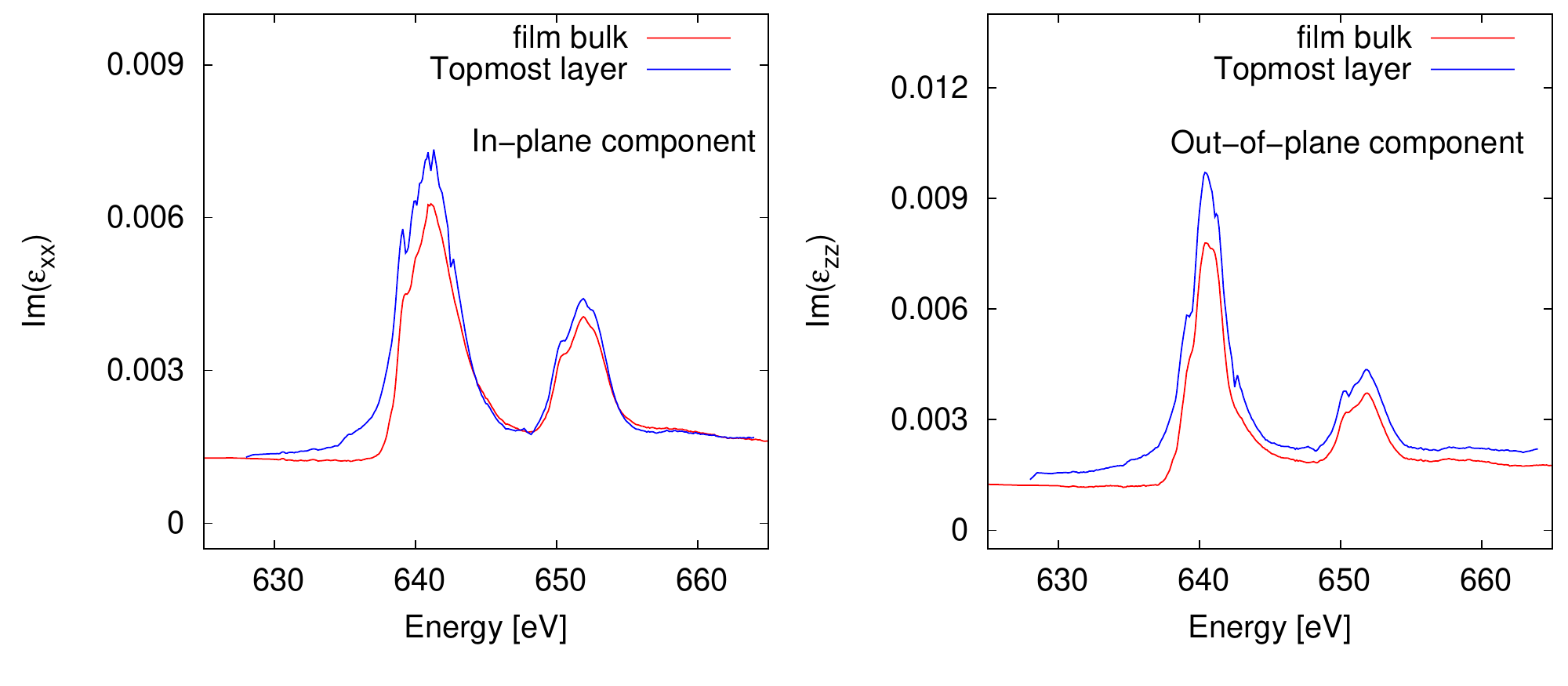}
\caption{The Kramers-Kronig constrained variational fit of the topmost MnO$_2$ layer reveals an electronic reconstruction at this interface. The spectral shape changes are caused by a local chemical environment that is different from the film bulk. Comparison with \cite{Choudhury2010} indicates that the increased weight at the L$_3$ edge might be explained by a reduction of the Mn valency.\\
The reflectivity for this fit is shown in Fig. \ref{fig:final-res}. The variational fit reduced $\chi^2$ to 0.22, which is an improvement of 32$\%$ over the best purely stoichiometric model.}
\label{fig:reconstruction}
\end{figure}

We searched for changes in the dielectric function at the top and bottom of the LSMO layer, as well as at in the transition layer. In each case, the stoichiometric parameters which we already introduced earlier and the dielectric function in-plane and out-of-plane were fitted simultaneously. We achieved the strongest improvement of the fit result when the variational fit approach was applied to the top of the LSMO film and when the changes of the dielectric constant were restricted to the topmost MnO$_2$ layer. More specifically, the $\chi^2$-value decreased from 0.33 to 0.22, showing the good agreement between model and experiment.
The result for the Mn L$_{2,3}$ spectrum of the topmost MnO$_2$ layer is shown in Fig.  \ref{fig:reconstruction}.
As can be observed in the figure, the lineshape of the topmost MnO$_2$-layer is different from the corresponding layer in the film bulk, thus implying a change in electronic properties of Mn. This can have multiple reasons. A reduction of the Mn valency from Mn$^{3+}$ to Mn$^{+(3-\delta)}$ for instance may occur as it has been reported for similar systems \cite{Valencia2006}. In addition, changes of chemical environment and local breaking of the symmetry are also common factors that induce changes in spectral lineshapes. A detailed analysis of the obtained lineshapes requires a detailed theoretical analysis, at least in terms of local multiplets models\,\cite{Mirone06}, which is beyond the scope of the present study and will be addressed in future work. The important result here is that our approach indeed yields depth-resolved  Mn L$_{2,3}$ spectra with atomic layer resolution.

\begin{figure}
\centering

\includegraphics[width=.95\columnwidth]{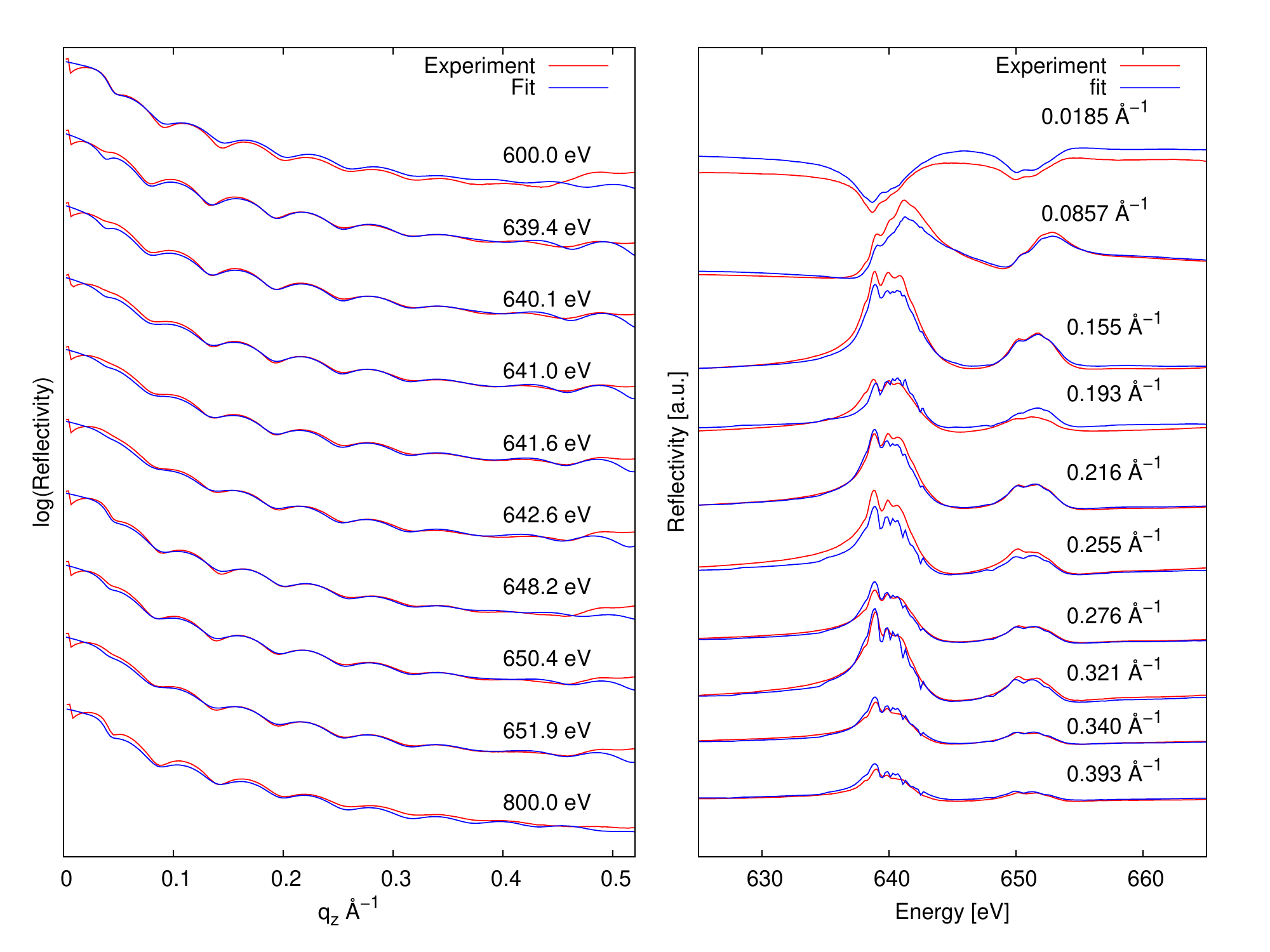}
\includegraphics[width=.95\columnwidth]{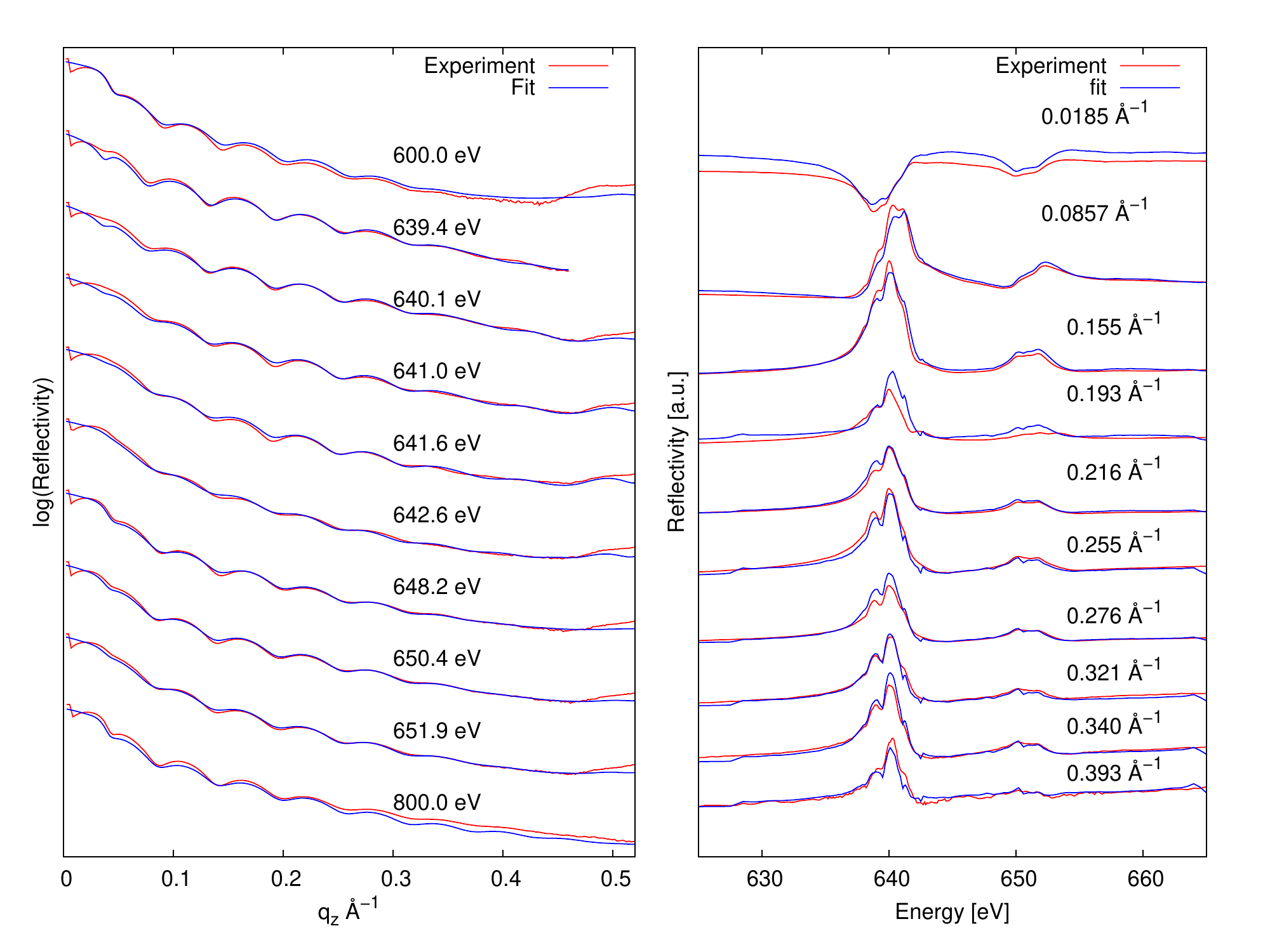}
\caption{Final fit result for the measured $\sigma$-polarized (top) and $\pi$-polarized (bottom) reflectivities, which includes the variational fit result of the topmost MnO$_2$ layer. Energy dependent scans were scaled to allow a good comparison.}
\label{fig:final-res}
\end{figure}

Figure \ref{fig:final-res} shows all the experimental reflectivities for both polarizations, together with the fit obtained with this final model. 
Although there are some deviations, the overall agreement is very good. Quite remarkable is that the polarization dependence is nicely captured. 

\section{Summary}
\label{sec:summary}

We have shown that the atomic structure of a material can influence soft x-ray RXR profiles and hence can be very important for the analysis of RXR data. 
A new approach for analyzing RXR in the soft x-ray range, which takes into account the atomic structure of a material, has been developed and applied to the RXR-analysis of an \lsmo~film.
The presented modeling in terms of atomic slices not only provides an improved description of the experimental data. It also enables to extract important additional information like layer termination and stacking sequence of the atomic planes of the film. Furthermore, it allows for a reliable extraction of spectral information about a specific layer of atoms inside a thin film. This additional information is essential for the understanding of  novel electronic phenomena generated at the interfaces and, hence,  renders RXR an even more powerful experimental tool to investigate artificial heterostructures and devices.\\

\section{Acknowledgements}
We thank M. Haverkort and U. Treske for fruitful discussions. M. Zwiebler, J. E. Hamann-Borrero and J. Geck gratefully acknowledge the support through the DFG Emmy Noether Program (Grants GE-1647/2-1 and HA6470/1-1).  Experiments described in this paper were performed at the Canadian Light Source, which is funded by the CFI, NSERC, NRC, CIHR, the Government of Saskatchewan, WD Canada and the University of Saskatchewan.

\section*{References}
\bibliographystyle{unsrt}
\bibliography{lsmo4_refl}

\begin{thebibliography}{10}

\bibitem{Dagotto07}
Elbio Dagotto.
\newblock When oxides meet face to face.
\newblock {\em Science}, 318(5853):1076--1077, 2007.

\bibitem{zubko11}
Pavlo Zubko, Stefano Gariglio, Marc Gabay, Philippe Ghosez, and Jean-Marc
  Triscone.
\newblock Interface physics in complex oxide heterostructures.
\newblock {\em Annu. Rev. Cond. Mat. Phys.}, 2(1):141--165, 2011.

\bibitem{Chakhalian12}
J.~Chakhalian, A.~J. Millis, and J.~Rondinelli.
\newblock Whither the oxide interface.
\newblock {\em Nat. Mater.}, 11:92, 2012.

\bibitem{hwang12}
H.~Y. Hwang, Y.~Iwasa, M.~Kawasaki, B.~Keimer, N.~Nagaosa, and Y.~Tokura.
\newblock Emergent phenomena at oxide interfaces.
\newblock {\em Nat. Mater.}, 11:103, 2012.

\bibitem{ohtomo04}
A.~Ohtomo and H.~Y. Hwang.
\newblock A high-mobility electron gas at the laalo$_3$/srtio$_3$
  heterointerface.
\newblock {\em Nature}, 427:423, 2004.

\bibitem{Buzdin05}
A.~I. Buzdin.
\newblock Proximity effects in superconductor-ferromagnet heterostructures.
\newblock {\em Rev. Mod. Phys.}, 77(3):935--976, Sep 2005.

\bibitem{Chakhalian06}
J.~Chakhalian, G.~Freeland, J. W.and~Srajer, J.~Strempfer, G.~Khaliullin, J.~C.
  Cezar, T.~Charlton, R.~Dalgliesh, C.~Bernhard, G.~Cristiani, H.-U.
  Habermeier, and B.~Keimer.
\newblock Magnetism at the interface between ferromagnetic and superconducting
  oxides.
\newblock {\em Nat Phys}, 2:244, 2006.

\bibitem{Chakhalian07}
J.~Chakhalian, J.~W. Freeland, H.-U. Habermeier, G.~Cristiani, G.~Khaliullin,
  M.~van Veenendaal, and B.~Keimer.
\newblock Orbital reconstruction and covalent bonding at an oxide interface.
\newblock {\em Science}, 318(5853):1114--1117, 2007.

\bibitem{eva11}
Eva Benckiser, Maurits~W. Haverkort, Sebastian Br\"uck, Eberhard Goering,
  Sebastian Macke, Alex Frañó, Xiaoping Yang, Ole~K. Andersen, Georg Cristiani,
  Hanns-Ulrich Habermeier, Alexander~V. Boris, Ioannis Zegkinoglou, Peter
  Wochner, Heon-Jung Kim, Vladimir Hinkov, and Bernhard Keimer.
\newblock Orbital reflectometry of oxide heterostructures.
\newblock {\em Nat. Mater.}, 10:189--193, 2011.

\bibitem{hamann14}
J.~E. Hamann-Borrero.
\newblock unpublished.

\bibitem{brueck11}
S~Br\"uck, S~Treiber, S~Macke, P~Audehm, G~Christiani, S~Soltan, H-U
  Habermeier, E~Goering, and J~Albrecht.
\newblock The temperature-dependent magnetization profile across an epitaxial
  bilayer of ferromagnetic la$_{2/3}$ca$_{1/3}$mno$_3$ and superconducting
  yba$_2$cu$_3$o$_{7-\delta}$.
\newblock {\em New J. Phys.}, 13(3):033023, 2011.

\bibitem{Tonnerre08}
J.~M. Tonnerre, M.~De~Santis, S.~Grenier, H.~C.~N. Tolentino, V.~Langlais,
  E.~Bontempi, M.~Garc\'ia-Fern\'andez, and U.~Staub.
\newblock Depth magnetization profile of a perpendicular exchange coupled
  system by soft-x-ray resonant magnetic reflectivity.
\newblock {\em Phys. Rev. Lett.}, 100:157202, Apr 2008.

\bibitem{Brueck10}
Sebastian Br\"uck, Sebastian Macke, Eberhard Goering, Xiaosong Ji, Qingfeng
  Zhan, and Kannan~M. Krishnan.
\newblock Coupling of fe and uncompensated mn moments in exchange-biased
  fe/mnpd.
\newblock {\em Phys. Rev. B}, 81:134414, Apr 2010.

\bibitem{Brown08}
S.~D. Brown, L.~Bouchenoire, P.~Thompson, R.~Springell, A.~Mirone, W.~G.
  Stirling, A.~Beesley, M.~F. Thomas, R.~C.~C. Ward, M.~R. Wells, S.~Langridge,
  S.~W. Zochowski, and G.~H. Lander.
\newblock Profile of the u $5f$ magnetization in $\mathrm{U}-\mathrm{Fe}$
  multilayers.
\newblock {\em Phys. Rev. B}, 77:014427, Jan 2008.

\bibitem{tonnerre12}
J.M. Tonnerre, E.~Jal, E.~Bontempi, N.~Jaouen, M.~Elzo, S.~Grenier,
  HL~Meyerheim, and M.~Przybylski.
\newblock Depth-resolved magnetization distribution in ultra thin films by soft
  x-ray resonant magnetic reflectivity.
\newblock {\em Eur. Phys. J. Special Topics}, 208:177--187, 2012.

\bibitem{Jal13}
Emmanuelle Jal, Maciej Dabrowski, Jean-Marc Tonnerre, Marek Przybylski,
  St\'ephane Grenier, Nicolas Jaouen, and J\"urgen Kirschner.
\newblock Magnetization profile across au-covered bcc fe films grown on a
  vicinal surface of ag(001) as seen by x-ray resonant magnetic reflectivity.
\newblock {\em Phys. Rev. B}, 87:224418, Jun 2013.

\bibitem{Hosoito14}
Nobuyoshi Hosoito, Takuo Ohkochi, Kenji Kodama, and Motohiro Suzuki.
\newblock Charge and induced magnetic structures of au layers in fe/au bilayer
  and fe/au/fe trilayer films by resonant x-ray magnetic reflectivity at the au
  $l_3$ absorption edge.
\newblock {\em J. Phys. Soc. Jpn.}, 83(2):024704, 2014.

\bibitem{Parrat54}
L.~G. Parratt.
\newblock Surface studies of solids by total reflection of x-rays.
\newblock {\em Phys. Rev.}, 95:359--369, Jul 1954.

\bibitem{Berreman72}
Dwight~W. Berreman.
\newblock Optics in stratified and anisotropic media: 4 $\times$ 4-matrix
  formulation.
\newblock {\em J. Opt. Soc. Am.}, 62(4):502--510, Apr 1972.

\bibitem{Ghebouli11}
M.A. Ghebouli, B.~Ghebouli, A.~Bouhemadou, M.~Fatmi, and K.~Bouamama.
\newblock Structural, electronic, optical and thermodynamic properties of
  sr$_x$ca$_{1-x}$o, ba$_x$sr$_{1-x}$o and ba$_x$ca$_{1-x}$o alloys.
\newblock {\em Journal of Alloys and Compounds}, 509(5):1440 -- 1447, 2011.

\bibitem{Vafaee13}
Mehran Vafaee, Mehrdad Baghaie~Yazdi, Aldin Radetinac, Gennady Cherkashinin,
  Philipp Komissinskiy, and Lambert Alff.
\newblock Strain engineering in epitaxial la$_{1-x}$sr$_{1+x}$mno$_4$ thin
  films.
\newblock {\em J. Appl. Phys.}, 113(5):053906, 2013.

\bibitem{Macke14}
Sebastian Macke, Abdullah Radi, Jorge~E. Hamann-Borrero, Martin Bluschke,
  Sebastian Br\"uck, Eberhard Goering, Ronny Sutarto, Feizhou He, Georg
  Cristiani, Meng Wu, Eva Benckiser, Hanns-Ulrich Habermeier, Gennady Logvenov,
  Nicolas Gauquelin, Gianluigi~A. Botton, Adam~P. Kajdos, Susanne Stemmer,
  Georg~A. Sawatzky, Maurits~W. Haverkort, Bernhard Keimer, and Vladimir
  Hinkov.
\newblock Element specific monolayer depth profiling.
\newblock {\em Adv. Mater.}, pages n/a--n/a, 2014.

\bibitem{nielsen11}
Jens Als-Nielsen and Des McMorrow.
\newblock {\em Elements of modern x-ray physics}.
\newblock Wiley, Chichester, England, 2011.

\bibitem{Lu2007}
Zhao Lu.
\newblock Accurate and efficient calculation of light propagation in
  one-dimensional inhomogeneous anisotropic media through extrapolation.
\newblock {\em J. Opt. Soc. Am. A}, 24(1):236--242, Jan 2007.

\bibitem{hawthorn11}
D.~G. Hawthorn, F.~He, L.~Venema, H.~Davis, A.~J. Achkar, J.~Zhang, R.~Sutarto,
  H.~Wadati, A.~Radi, T.~Wilson, G.~Wright, K.~M. Shen, J.~Geck, H.~Zhang,
  V.~Nov\'{a}k, and G.~A. Sawatzky.
\newblock An in-vacuum diffractometer for resonant elastic soft x-ray
  scattering.
\newblock {\em Rev. Sci. Instrum.}, 82(7):073104, 2011.

\bibitem{chantler71}
C.~T. Chantler.
\newblock Theoretical form factor, attenuation, and scattering tabulation for
  z=1--92 from e=1--10 ev to e=0.4--1.0 mev.
\newblock {\em J. Phys. Chem. Ref. Data}, 24(1):71--643, 1995.

\bibitem{Baumann2003}
C.~Baumann, G.~Allodi, B.~Buchner, R.~De Renzi, P.~Reutler, and
  A.~Revcolevschi.
\newblock Magnetism of la1-xsr1+xmno4 as revealed by µsr.
\newblock {\em Physica B: Condensed Matter}, 326(1-4):505 -- 508, 2003.

\bibitem{haverkort-PRB2010}
M.~W. Haverkort, N.~Hollmann, I.~P. Krug, and A.~Tanaka.
\newblock Symmetry analysis of magneto-optical effects: The case of x-ray
  diffraction and x-ray absorption at the transition metal $l_{2,3}$ edge.
\newblock {\em Phys. Rev. B}, 82(9):094403, Sep 2010.

\bibitem{Macke2014}
Sebastian Macke, Abdullah Radi, Jorge~E. Hamann-Borrero, Adriano Verna, Martin
  Bluschke, Sebastian Brück, Eberhard Goering, Ronny Sutarto, Feizhou He, Georg
  Cristiani, Meng Wu, Eva Benckiser, Hanns-Ulrich Habermeier, Gennady Logvenov,
  Nicolas Gauquelin, Gianluigi~A. Botton, Adam~P. Kajdos, Susanne Stemmer,
  Georg~A. Sawatzky, Maurits~W. Haverkort, Bernhard Keimer, and Vladimir
  Hinkov.
\newblock Element specific monolayer depth profiling.
\newblock {\em Advanced Materials}, pages n/a--n/a, 2014.

\bibitem{Bertacco02}
R~Bertacco, J.P Contour, A~Barthélemy, and J~Olivier.
\newblock Evidence for strontium segregation in la$_{0.7}$sr$_{0.3}$mno$_3$
  thin films grown by pulsed laser deposition: consequences for tunnelling
  junctions.
\newblock {\em Surface Science}, 511(1-3):366 -- 372, 2002.

\bibitem{Fister08}
Tim~T. Fister, Dillon~D. Fong, Jeffrey~A. Eastman, Peter~M. Baldo, Matthew~J.
  Highland, Paul~H. Fuoss, Kavaipatti~R. Balasubramaniam, Joanna~C. Meador, and
  Paul~A. Salvador.
\newblock In situ characterization of strontium surface segregation in
  epitaxial la$_{0.7}$sr$_{0.3}$mno$_3$ thin films as a function of oxygen
  partial pressure.
\newblock {\em Appl. Phys. Lett.}, 93(15):151904, 2008.

\bibitem{Katsiev09}
Khabiboulakh Katsiev, Bilge Yildiz, Kavaipatti Balasubramaniam, and Paul~A.
  Salvador.
\newblock Electron tunneling characteristics on la$_{0.7}$sr$_{0.3}$mno$_3$
  thin-film surfaces at high temperature.
\newblock {\em Appl. Phys. Lett.}, 95(9):--, 2009.

\bibitem{Abellan11}
P.~Abellán, C.~Moreno, F.~Sandiumenge, X.~Obradors, and M.-J. Casanove.
\newblock Misfit relaxation of la$_{0.7}$sr$_{0.3}$mno$_3$ thin films by a
  nanodot segregation mechanism.
\newblock {\em Appl. Phys. Lett.}, 98(4):--, 2011.

\bibitem{Li12}
Zhipeng Li, Michel Bosman, Zhen Yang, Peng Ren, Lan Wang, Liang Cao, Xiaojiang
  Yu, Chang Ke, Mark B.~H. Breese, Andrivo Rusydi, Weiguang Zhu, Zhili Dong,
  and Yong~Lim Foo.
\newblock Interface and surface cation stoichiometry modified by oxygen
  vacancies in epitaxial manganite films.
\newblock {\em Adv. Funct. Mater.}, 22(20):4312--4321, 2012.

\bibitem{Poggini2014}
L.~Poggini, S.~Ninova, P.~Graziosi, M.~Mannini, V.~Lanzilotto, B.~Cortigiani,
  L.~Malavolti, F.~Borgatti, U.~Bardi, F.~Totti, I.~Bergenti, V.~A. Dediu, and
  R.~Sessoli.
\newblock A combined ion scattering, photoemission, and dft investigation on
  the termination layer of a la0.7sr0.3mno3 spin injecting electrode.
\newblock {\em The Journal of Physical Chemistry C}, 118(25):13631--13637,
  2014.

\bibitem{Haage97}
T.~Haage, J.~Zegenhagen, J.~Q. Li, H.-U. Habermeier, M.~Cardona, Ch. Jooss,
  R.~Warthmann, A.~Forkl, and H.~Kronm\"uller.
\newblock Transport properties and flux pinning by self-organization in
  yba$_2$cu$_3$o$_{7-\delta}$ films on vicinal srtio$_3$(001).
\newblock {\em Phys. Rev. B}, 56:8404--8418, Oct 1997.

\bibitem{Malik12}
V.~K. Malik, I.~Marozau, S.~Das, B.~Doggett, D.~K. Satapathy, M.~A.
  Uribe-Laverde, N.~Biskup, M.~Varela, C.~W. Schneider, C.~Marcelot, J.~Stahn,
  and C.~Bernhard.
\newblock Pulsed laser deposition growth of heteroepitaxial
  yba$_2$cu$_3$o$_7$/la$_{0.67}$ca$_{0.33}$mno$_3$ superlattices on ndgao$_3$
  and sr$_{0.7}$la$_{0.3}$al$_{0.65}$ta$_{0.35}$o$_3$ substrates.
\newblock {\em Phys. Rev. B}, 85:054514, Feb 2012.

\bibitem{nevot80}
L.~Nevot and P.~Croce.
\newblock Caract{\'e}risation des surfaces par r{\'e}flexion rasante de rayons
  x. application {\`a} l'{\'e}tude du polissage de quelques verres silicates.
\newblock {\em Revue de Physique appliqu{\'e}e}, 15(3):761--779, 1980.

\bibitem{Kuzmenko2005}
A.~B. Kuzmenko.
\newblock Kramers-kronig constrained variational analysis of optical spectra.
\newblock {\em Review of Scientific Instruments}, 76(8):--, 2005.

\bibitem{Stone2012}
Kevin~H. Stone, S.~Manuel Valvidares, and Jeffrey~B. Kortright.
\newblock Kramers-kronig constrained modeling of soft x-ray reflectivity
  spectra: Obtaining depth resolution of electronic and chemical structure.
\newblock {\em Phys. Rev. B}, 86:024102, Jul 2012.

\bibitem{Valencia2006}
S.~Valencia, A.~Gaupp, W.~Gudat, Ll. Abad, Ll. Balcells, A.~Cavallaro,
  B.~Martinez, and F.~J. Palomares.
\newblock Mn valence instability in la$_{2/3}$ca$_{1/3}$mno$_3$ thin films.
\newblock {\em Phys. Rev. B}, 73:104402, Mar 2006.

\bibitem{Mirone06}
A.~Mirone, S.~S. Dhesi, and G.~van~der Laan.
\newblock Spectroscopy of la0.5sr1.5mno4 orbital ordering: a cluster many-body
  calculation.
\newblock {\em Eur. Phys. J. B}, 53(1):23--28, 2006.

\bibitem{Choudhury2010}
Sanjukta Choudhury.
\newblock Spectroscopic study of transition metal compounds.
\newblock {\em University of Saskatchewan Electronic Theses and Dissertations},
  2010.

\end{thebibliography}

\end{document}